\documentclass{mn2e}
\usepackage{times}
\usepackage{graphicx}
\usepackage{soul}       

\def\Msolar{\hbox{${\rm M}_\odot$}}

\begin{document} 

\title[Interstellar extinction near 30\,Doradus]{Probing
interstellar extinction near the 30\,Doradus nebula with red giant
stars\thanks{Based on observations with the NASA/ESA
{\it Hubble Space Telescope}, obtained at the Space Telescope Science
Institute, which is operated by AURA, Inc., under NASA contract
NAS5-26555.}}

\author[Guido De Marchi, Nino Panagia, L\'eo Girardi]
{Guido~De~Marchi,$^1$ Nino~Panagia,$^{2,3,4}$ and L\'eo~Girardi$^5$\\
$^1$European Space Research and Technology Centre, Keplerlaan 1, 2200 AG
Noordwijk, The Netherlands, gdemarchi@rssd.esa.int \\
$^2$Space Telescope Science Institute, 3700 San Martin Drive, Baltimore, MD
21218, USA, panagia@stsci.edu\\
$^3$INAF--NA, Osservatorio Astronomico di Capodimonte, Salita Moiariello
16, 80131 Naples, Italy\\
$^4$Supernova Ltd, OYV \#131, Northsound Rd., Virgin Gorda VG1150,
Virgin Islands, UK\\
$^5$INAF--PD, Osservatorio Astronomico di Padova, Vicolo Osservatorio 5,
35122 Padua, Italy, leo.girardi@oapd.inaf.it}

\date{Received 9.10.2013; Accepted 13.11.2013}
\pagerange{\pageref{firstpage}--\pageref{lastpage}} \pubyear{2013}

\maketitle

\begin{abstract}

We have studied the interstellar extinction in a field of $\sim 3\arcmin
\times 3\arcmin$ located about $6\arcmin$ SW of 30\,Doradus in the Large
Magellanic Cloud (LMC). {\em Hubble Space Telescope} observations in the
$U$, $B$, $V$, $I$ and $H\alpha$ bands reveal patchy extinction in this
field. The colour--magnitude diagram (CMD) shows an elongated stellar
sequence, almost parallel to the main sequence (MS), which is in reality
made up of stars of the red giant clump (RC) spread across the CMD by
the uneven levels of extinction in this region. Since these objects are
all at the same distance from us and share very similar physical
properties, we can derive quantitatively both the extinction law in the
range $3\,000-8\,000$\,\AA\ and the absolute extinction towards about
100 objects, setting statistically significant constraints on the dust
grains properties in this area. We find an extinction curve considerably
flatter than the standard Galactic one and than those obtained before
for the LMC. The derived value of $R_V=5.6\pm0.3$ implies that in this
region larger grains dominate. Upper MS stars span a narrower range of
$E(B-V)$ values than RC objects, at variance with what has been found
elsewhere in the LMC.

\end{abstract}


\section{Introduction}

The traditional approach to deriving extinction curves is the ``pair
method,'' in which the {flux distribution} or colours of a reddened
object are compared with those of a star of the same spectral type (e.g.
Massa, Savage \& Fitzpatrick 1983; Cardelli et al. 1992; and references
therein). This is possible because the interstellar extinction
$A_\lambda$ towards a given star can be determined by the ratio of its
observed spectral energy distribution $F_{\rm obs}(\lambda)$ and the
expected or intrinsic spectrum for this star $F_{\rm exp}(\lambda)$,
i.e.

\begin{equation}
A_\lambda = -2.5 \, \times \, \log \frac{F_{\rm obs}(\lambda)}
{F_{\rm exp}(\lambda)}
\label{eq1}
\end{equation}

\noindent 
In principle, the expected spectrum should be the one appropriate for an
extinction-free star with the same spectral type, the same absolute
luminosity and the same distance as the target star. Since in general it
is very hard to satisfy these conditions accurately, most of the studies
are done considering the colour excess, $E_\lambda (\lambda,
\lambda_0)$, defined as

\begin{equation} 
E_\lambda (\lambda, \lambda_0) = A_\lambda - A_{\lambda_0} = 
-2.5 \,\times\, \log \left[\frac{F_\lambda / 
F_{\lambda_0}}{F_{\lambda,{\rm ref}}/F_{\lambda_0,{\rm ref}}}\right]
\label{eq2} 
\end{equation}

\noindent
Dealing with a ratio of ratios, this quantity is independent of the
distance to the stars considered as well as of their radii. Moreover, to
separate the behaviour of the extinction from its absolute value for
different pairs of stars or, equivalently, for different lines of sight,
the colour excess is usually normalized. It is customary to take the
ratio between a colour excess and the one in a reference wavelength
interval $(\lambda_1, \lambda_0)$, i.e.

\begin{equation}
\mathcal{E}(\lambda, \lambda_0) =
\frac{E(\lambda,\lambda_0)}{E(\lambda_1,\lambda_0)},
\label{eq3}
\end{equation}

\noindent 
which for example for $\lambda_1=\lambda_B\simeq 4\,350$\,\AA, and
$\lambda_0=\lambda_V \simeq 5\,500$\,\AA\ becomes $E(\lambda -
V)/E(B-V)$. The colour excess normalized in this way is the form usually
adopted to describe the extinction curve, i.e. the run of the selective
extinction, suitably normalized, as a function of the wavelength
$\lambda$. 

{We note that an implicit assumption underlying this procedure is
that the extinction law is the same for all pairs of stars.} With this
procedure we can study and compare the slopes of the extinction laws
between different stars and directions, and therefore derive relative
dust properties, but still we do not have a {characterization} of
the absolute value of the extinction. The latter is usually expressed in
terms of the absolute extinction  $A_\lambda/A_V$ at wavelength
$\lambda$, comparing the total extinction at that wavelength with that
in the $V$ band. Alternatively, it is also given by the parameter $R_V
\equiv A_V/E(B-V)$, which provides total to selective extinction in the
$B$ and $V$ bands. 

To derive the absolute extinction measurements have to be extended to 
very long wavelengths, taking advantage of the physical properties of
dust grains, whose extinction cross-section tends to zero when the
wavelength approaches infinity, i.e. $\lim_{\lambda \to \infty}
A_\lambda = 0$, and therefore the colour excess
$\mathcal{E}(\lambda,\lambda_0)$ becomes

\begin{equation}
\mathcal{E}(\lambda, \infty) = \frac{A_\lambda- A_\infty}{A_B - A_V} =
\frac{A_\lambda}{E(B-V)}.
\label{eq4}
\end{equation}

For many lines of sight accurate measurements today extend to the 
near-infrared, usually to the $K$ band, and it is a common practice to
estimate values for $R_V$ from the extrapolation 

\begin{equation}
R_V = \frac{E(V-\lambda_\infty)}{E(B-V)} \simeq \, 1.1 \, 
      \frac{E(V - K)}{E(B - V)}
\label{eq5}
\end{equation}

\noindent 
based on the assumption that the theoretical extinction curve no. 15 of 
van de Hulst (van de Hulst 1949, as reported by Johnson 1968) is
appropriate for the interstellar medium under study. However, as
Fitzpatrick \& Massa (2007) already pointed out, this extrapolation is
necessarily only accurate if $R_V \simeq 3$, since this is the value to 
which the van de Hulst's theoretical curve no. 15 applies, but it
systematically deviates at higher and lower values of $R_V$.

That $R_V$ varies considerably in our Galaxy is well known, particularly
thanks to the studies conducted in the 1970's with the {\em Orbiting
Astronomical Observatory 2} (Bless \& Savage 1972), with the {\em
Copernicus} and {\em TD--1} satellites (e.g. Seaton 1979), and later
with the {\em International Ultraviolet Explorer}, which revealed a wide
variety of extinction curves (see e.g. Fitzpatrick 1998, 1999 and
references therein). Large variations are seen in the ultraviolet (UV) 
portion of the
spectrum, with a more or less steep raise of the curve in the far-UV and
a more or less prominent ``bump'' at 2\,175\,\AA, and in smaller amount
also at optical wavelengths.

Using the extrapolation to infinity mentioned above to derive $R_V$,
Gordon et al. (2003) have carried out a quantitative comparison of all
known extinction curves in the Large and Small Magellanic Clouds (LMC and
SMC, respectively) with the Galactic curves. They find average $R_V$
values of $2.05 \pm 0.17$ for the SMC wing (one star), $2.74 \pm 0.13$
for the SMC bar (four stars), $2.76 \pm 0.09$ for the LMC supershell
near the 30\,Dor region (eight stars) and $3.41\pm 0.06$ for lines of
sight to ten stars  elsewhere in the LMC. Although these $R_V$ values
are not too different from the average value found in the diffuse
Galactic interstellar medium ($R_V=3.1$; e.g. Savage \& Mathis 1979),
the majority of the Magellanic Clouds (MC) extinction curves that Gordon
et al. (2003) find are significantly different in the UV from the Milky
Way (MW) extinction curves. These authors conclude that the environments
probed by their observations, being biased towards active star forming
regions, are rather different from those included in the sample studied
by Cardelli et al. (1988, 1989), which is based on fairly quiescent
lines of sight.

Recently, Haschke, Grebel \& Duffau (2011), arguing that the differences
between the MC extinction curves of Gordon et al. (2003) and the
Galactic law by Cardelli et al. (1989)  are quite small at optical
wavelengths, concluded that the Cardelli et al. (1989) curve is
generally a valid assumption also for the MC in that wavelength range.
In fact, the differences appear small because the absolute extinction
$A_\lambda/A_V$ given by Gordon et al. (2003) is normalized in the $V$
band, but considerable differences between the MC and the Galaxy remain
also at these wavelengths.

In this work we present a new method to unambiguously determine the
absolute value of the extinction in the MC in all observed bands,
deriving in this way an absolute extinction law. Our method, originally
presented by Panagia \& De Marchi (2005), makes use of
multi-band photometry of red giant stars belonging to the red clump
(RC). 

Other authors have used RC observations before to study the reddening
distribution and to derive reddening maps in the MC (e.g. Zaritsky 1999;
Haschke et al. 2011; Tatton et al. 2013), for an assumed extinction
 curve. However, these observations do not provide an independent 
determination of the extinction curve in the MC. Recently, Nataf et al. 
(2013) used observations of RC stars in the Galactic bulge, collected 
as part of the Optical Gravitational Lensing Experiment (OGLE) project 
(Paczynski 1986), to determine the reddening and extinction towards the 
inner MW, finding a rather steep extinction curve ($R_V \simeq 2.5$). 
However, since only observations in the $V$ and $I$ bands are available 
as part of the OGLE project (supplemented by $E(J-K_s)$ measurements from 
other surveys), these results cannot be extended to shorter wavelengths. 

Instead, the novelty and advantage of our method consists in the facts
that {\em (i)} all stars on which we operate are at the same distance,
to better than 1\,\%, {\em (ii)} they have very similar intrinsic
physical properties in all bands, within $0.05$\,mag for similar age and
metallicity, {\em (iii)} our statistics is very solid, with of the order of
$\sim 100$ objects per field, and {\em (iv)} we derive a self-consistent
{\em absolute} extinction curve over the entire optical range ($3\,000 -
8\,000$\,\AA) from photometry alone, without uncertain extrapolations
from infrared bands. Our method, whose application is shown here for a
field in the vicinity of the 30\,Dor nebula in the LMC, can easily be
extended to other nearby galaxies.

The structure of this paper is as follows. The observations and their
analysis are presented in Section\,2, while the results of the
photometric reductions are shown in Section\,3. The physical properties
of the red giant clump are discussed in Section\,4, while in Section\,5
we proceed to deriving the absolute extinction law in the field of our
study. Section\,6 is devoted to a discussion of the reddening
distribution in this field. A summary and our conclusions follow in
Section\,7.

\vspace*{-0.5cm} 
\section{Observations and data analysis}

The data used in our investigation were collected as part of the
{\em Hubble Space Telescope (HST) / Wide Field Planetary Camera\,2} (WFPC\,2)
 pure parallel programme (see Wadadekar et al. 2006) and
cover a region of $\sim 2\farcm7$ on a side located at $\alpha=5^{\rm h}
37^{\rm m} 49^{\rm s}$, $\delta=-69^{\circ} 8^\prime 18^{\prime\prime}$
(J\,2000), or about $6\arcmin$\,SW of 30\,Dor (NGC\,2070). This region was
observed with the WFPC\,2 instrument (Mc Master \& Biretta 2008) on
board the HST on 2001 Jan 6 and 7 in several
bands. Table\,1 lists the total duration of the long exposures in each
band. In addition to these, a series of short exposures (typically $\sim
10$\,s duration) was collected in all bands to properly measure the
magnitude of bright stars that would otherwise saturate. 

\begin{table}
\centering 
\caption{Cumulative exposure times in the various bands}
\begin{tabular}{llc} 
\hline
\multicolumn{2}{l}{Filter} &  Total exposure time\\
\hline      
F300W & ($U$) & 2\,800 s\\
F450W & ($B$) & 2\,500 s\\
F606W & ($V$) & 2\,900 s\\
F656N & ($H\alpha$) & 3\,800 s\\
F814W & ($I$) & 3\,200 s\\ 
\hline
\end{tabular}
\label{tab1}
\end{table}

\begin{figure*}
\centering
\resizebox{\hsize}{!}{\includegraphics[width=16cm]{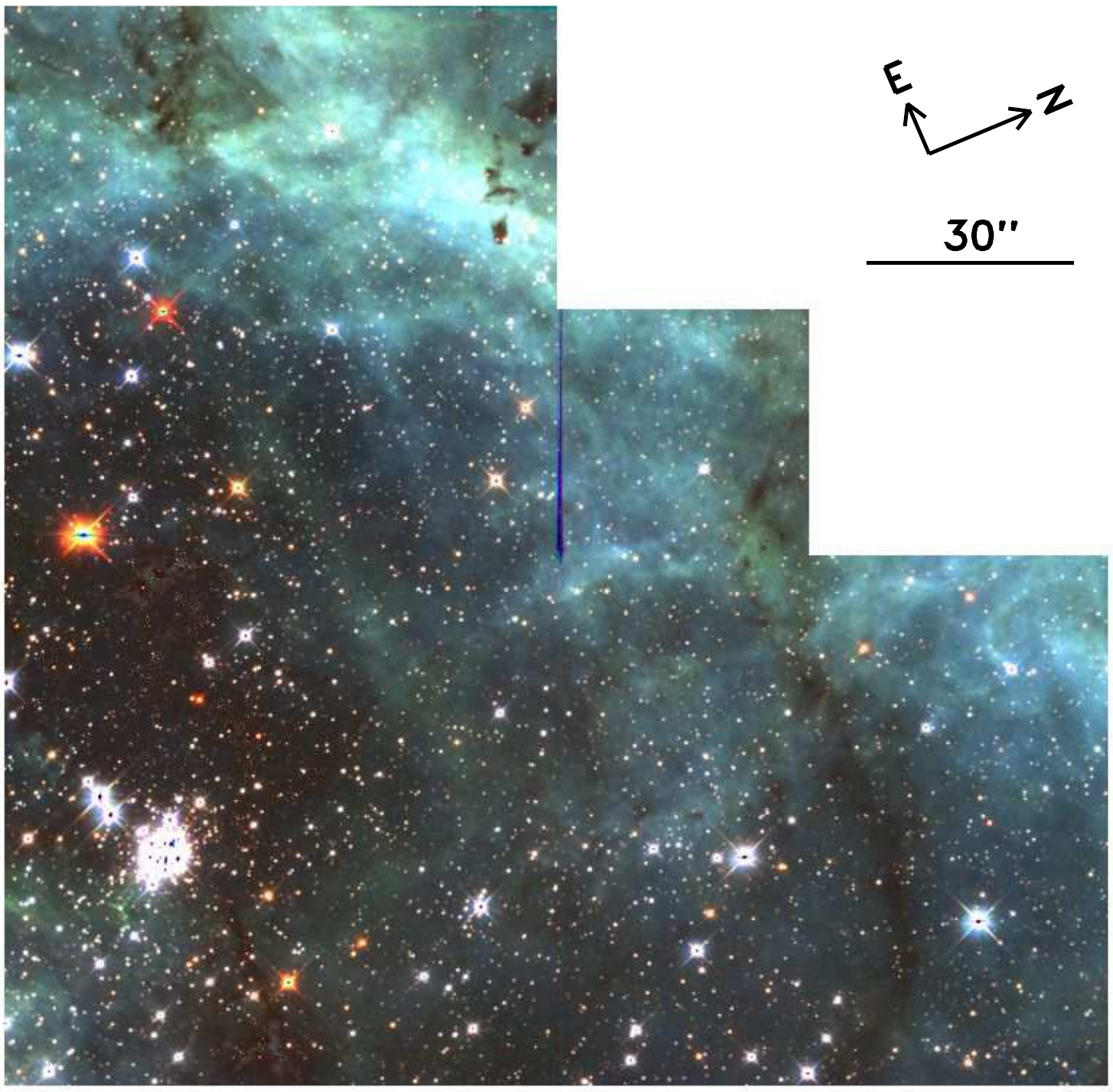}}
\caption{True colour image of a field of $2\farcm7 \times 2\farcm7$, 
located at  RA$=5^{\rm h} 37^{\rm m} 49^{\rm s}$, Dec$=-69^{\circ}
8^\prime 18^{\prime\prime}$ (J\,2000), or about $6\arcmin$\,SW of
30\,Dor. North is inclined $\sim 67^\circ$ clockwise with respect to the
vertical, and east is to the left of north. In this figure, the exposure
in the F300W band is used for the blue channel, the F450W image for the
green channel, and the average of the F606W and F814W exposures for the
red channel. In this work we only consider objects contained in the
chips WF\,2 (upper left), WF\,3 (lower left) and WF\,4 (lower right),
whereas stars in the PC\,1 chip (upper right) are not considered.}
\label{fig1}
\end{figure*}

A suitable dithering pattern was applied during the observations in
order to make it possible to efficiently remove cosmic rays by comparing
images taken through the same filter at different times. 

The data were subjected to the automated standard reduction and
pipeline procedure offered by the HST archive (``on the fly
calibration''; Micol et al. 2000), which also took care of properly
registering and co-adding images taken in the same band. The accuracy of
the registration and combination procedures was verified by comparing
the properties of the point spread function (PSF) in the individual images 
and in the combined ones. We find that the full width at half-maximum (FWHM) of stars in
the combined frame increases by about 15\,\%, as is typical of any
shift-and-add operation conducted on under-sampled images. 

In Figure\,\ref{fig1} we show a true-colour image obtained by combining
the F450W, F606W and F814W frames with the {\rm F656N} ($H\alpha$)
observations. The $H\alpha$ image traces the extended nebulosity, which
appears highly variable on scales of the order of $1\arcsec$ or $\sim 0.25$\,pc
at the distance of the LMC ($51.4 \pm 1.2$\,kpc; Panagia et al. 1991,
and updates in  Panagia 1998, 1999). Several obscured regions
are also present. It is, therefore, to be expected that the extinction
will vary considerably across this field.

\begin{figure*}
\centering
\resizebox{\hsize}{!}{\includegraphics[angle=90]{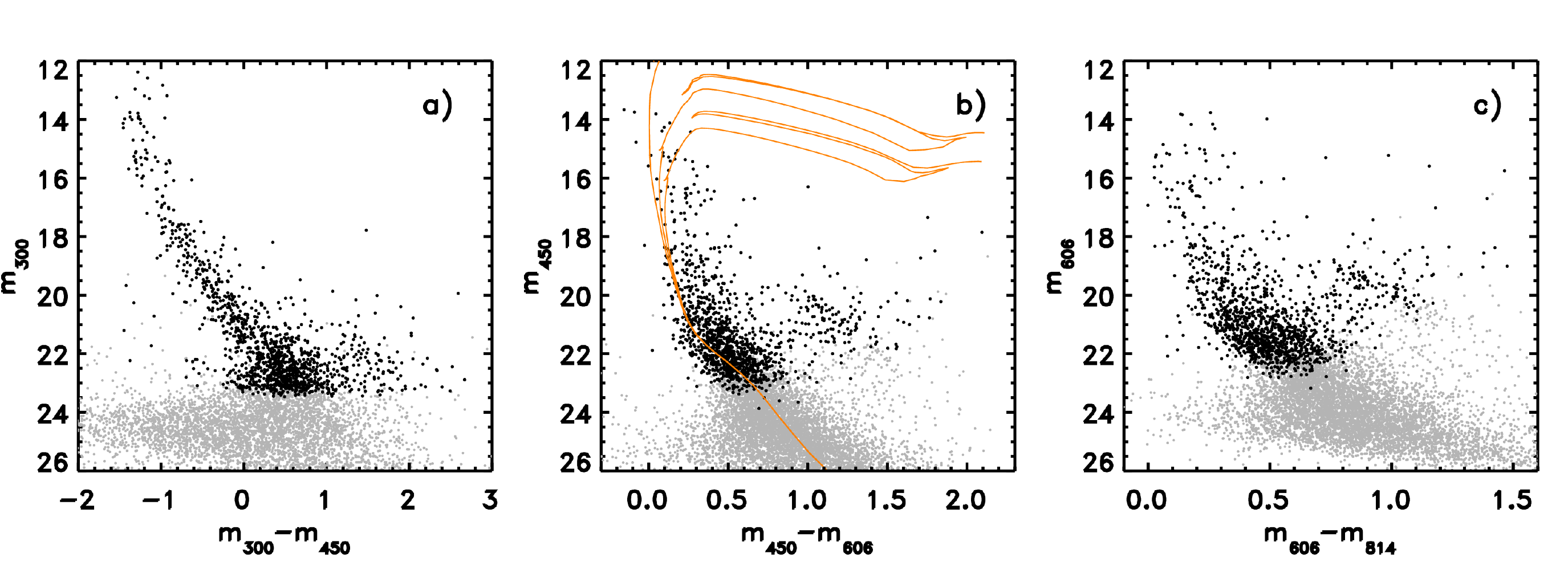}}
\caption{Colour--magnitude diagrams in the most common colour
combinations. Thicker dots correspond to stars with $\delta_4 < 0.1$.
{The isochrones shown in panel b) correspond, from left to right, to 
ages of 4, 20 and 40\,Myr for metallicity $Z=0.008$.}}
\label{fig2}
\end{figure*}

Object identification was carried out on the F814W combined frame, the
deepest of the set, using the standard {\sc IRAF} {\em apphot.daofind}
routine. To guarantee that most noise spikes and PSF artefacts are
automatically rejected, we conservatively set the detection threshold at
$10\,\sigma$ above the local average background level. The automated
procedure turned out about 8\,830 stars in the three wide field (WF) 
chips (we did
not use the PC frame), all of which were carefully inspected by eye to
remove  saturated stars, a number of features above $10\,\sigma$ (PSF
tendrils, noise spikes, etc.) that {\em daofind} had interpreted as
stars, as well as a very few extended objects (with FWHM larger than
twice that typical of point sources). The final object list pruned in
this way includes about 8\,350 well defined stars in the F814W band and
was used as a master list for photometry in all frames in all bands. 

Crowding not being severe, stellar fluxes were measured by using the
standard {\sc IRAF} {\it digiphot.apphot} aperture photometry routine,
following the prescription of the ``core aperture photometry'' technique
described in Gilmozzi \& Panagia (1991) and De Marchi et al. (1993). In
particular, we adopted an aperture radius of 2 pixel and a background
annulus extending from 3 to 5 pixel in radius. Aperture corrections were
calculated for an infinite aperture and the instrumental magnitudes
calibrated in the {\em HST} magnitude system (VEGAMAG) by adopting the 
zero-points listed in the WFPC\,2 Data Handbook (Gonzaga \& Biretta 2010).
Saturated stars were measured in the short exposures and the resulting
magnitudes properly rescaled for the exposure time. Owing to the
considerable inefficiency with which charges are transferred in the
WFPC\,2 detectors, the magnitude of a star depends on its position on
the detector and must, therefore, be corrected accordingly. To this aim,
we have followed the prescriptions of Dolphin  (2000).

Of the originally identified $\sim 8\,350$ stars in F814W, 78\,\% have a
photometric uncertainty smaller than $0.1$ mag. The fraction of stars
with uncertainty on the magnitudes smaller than $0.1$ mag is 67\,\% in
F606W, 37\,\% in F450W, 7\,\% in F300W and 6\,\% in {F656N}.
Following Romaniello et al. (2002), it is convenient to define the mean
error of each star in the four broad-bands ($\delta_4$) as:

\begin{equation}
\delta_4 = \sqrt{\frac{\delta^2_{300} + \delta^2_{450} +
                       \delta^2_{606} + \delta^2_{814}}{4}}
\label{eq6}
\end{equation}

\noindent    
where $\delta_{300}$, $\delta_{450}$, $\delta_{606}$ and $\delta_{814}$
are the uncertainties in each individual band.\footnote{The definition
given by Equation\,\ref{eq6} can be generalized for any combination of
bands. For instance, in Section\,5 we will refer to the combined
photometric uncertainty $\delta_3$ in the $V$ (F606W), $I$ (F814W) and
$H\alpha$ (F656N) bands.} A total of $\sim 1\,300$ stars have a mean
error $\delta_4 < 0.1$ mag and they all have broad-band magnitudes
brighter than $m_{300} \simeq 23.5$, $m_{450} \simeq 23.9$, $m_{606}
\simeq 23.2$ and $m_{814} \simeq  22.5$. The photometric completeness at
these relatively bright magnitudes only depends on the stellar density
and the presence of bright saturated stars. Since the average
star-to-star distance at $m_{606} \simeq 23.2$ is $\sim 22$\,pixel and
the FWHM of the PSF is less than 2\,pixel, the photometric completeness
is very close to 100\,\%.

\vspace*{-0.3cm}
\section{Colour--magnitude diagrams}

In Figure\,\ref{fig2} we show the colour--magnitude diagrams (CMDs)
obtained by combining the data of all three WF chips in the broad-band
filters. We show the CMDs corresponding to the most common colour
combinations. Objects with an average photometric error $\delta_4 < 0.1$
are marked with thicker and darker dots. As indicated above, the
inspection of Figure\,\ref{fig1} already suggests the presence of patchy
absorption. This is further confirmed by the CMDs of Figure\,\ref{fig2},
where for instance the upper stellar main sequence (MS) appears, in all
colours, considerably broader than the photometric uncertainty typical
of stars with $\delta_4 < 0.1$\,mag. Some of the observed broadening can
be caused by an age spread across the field, particularly in the upper
MS (UMS) populated by B-type stars that evolve on a time-scale (few Myr)
comparable with that of the star formation process itself (e.g. Iben
1974). 

For reference, the lines plotted to the left of the MS in
Figure\,\ref{fig2}b correspond, from left to right, to isochrones for a
population with metallicity $Z=0.008$ and ages of, respectively, 4, 20
and 40\,Myr from the models  of Marigo et al. (2008) for the specific
WFPC\,2 bands used here. For purposes of illustration, the isochrones
have been reddened according  to the Galactic extinction law, by an
amount corresponding to $E(B-V)=0.3$. This value includes the
intervening MW absorption along the line of sight to the LMC, which
accounts for $E(B-V)=0.07$ or $A_V = 0.22$ (Fitzpatrick \& Savage 1984).
As for the  distance, throughout this paper we will assume the
distance modulus for the LMC as that of SN\,1987A determined by Panagia
(1998), namely $(m-M)_0=18.55$, corresponding to a distance of
$51.4$\,kpc as mentioned above. While some of the colour spread at the
brightest end can be caused by age, it appears that without a
considerable spread in reddening it would be hard to justify the
broadening observed at $m_{606} \la 20$, where all stars have
photometric uncertainty $\delta_4 < 0.1$.  

\begin{figure}
\centering
\resizebox{\hsize}{!}{\includegraphics[angle=90]{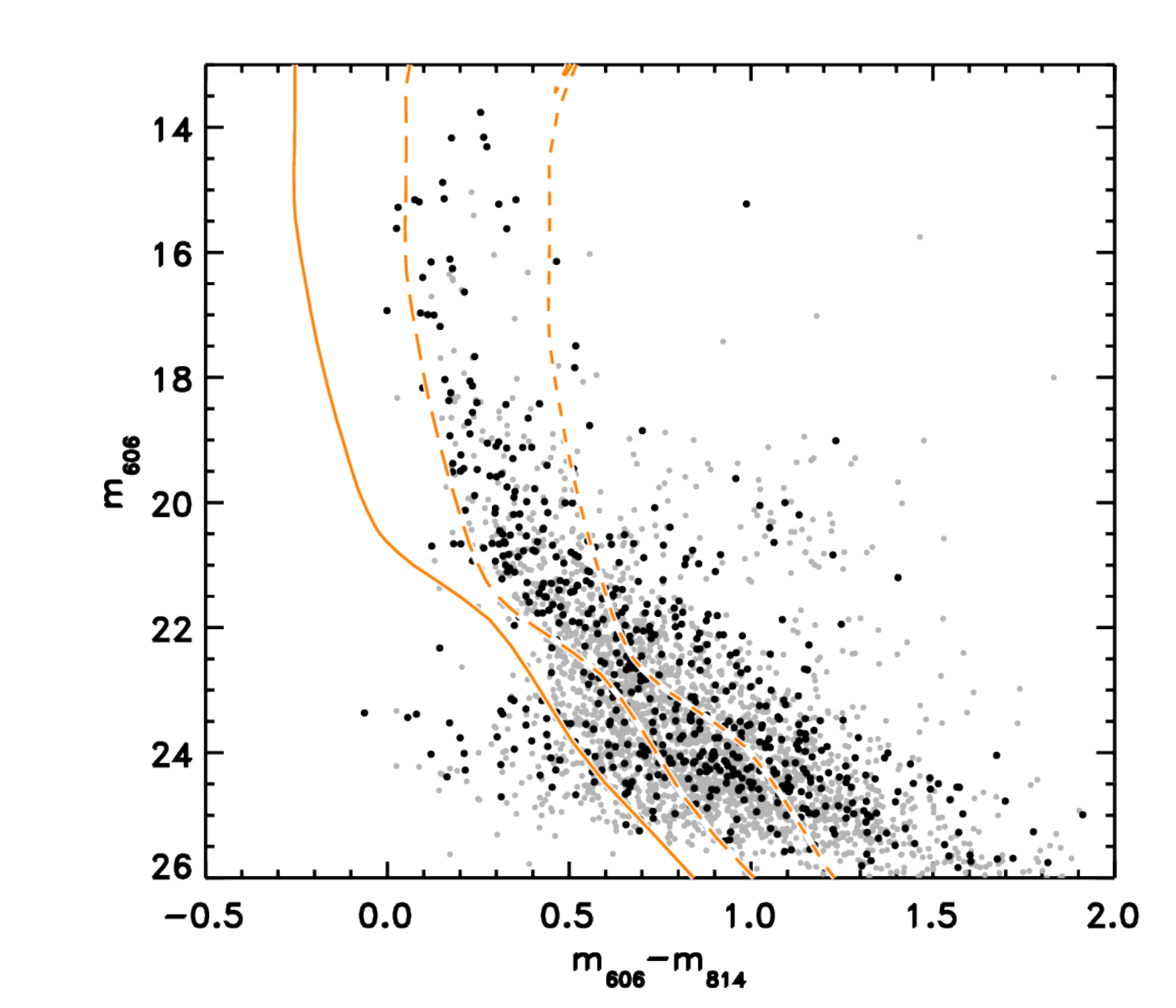}}
\caption{Same CMD as in Figure\,\ref{fig2}c but only for the stars
inside the WF\,3 chip. Darker dots are used for stars located within 3
times the half-light radius of the small cluster near the lower-left
corner of Figure\,\ref{fig1}. Light grey dots correspond to the rest of
the stars in WF\,3. Isochrones from Marigo et al. (2008) for age of 4\,Myr
and three values of $E(B-V)$,  namely 0, $0.3$, and $0.7$. }
\label{fig3}
\end{figure}

Furthermore, strong indication of the presence of differential reddening
on very small scales comes from the CMD of the stars belonging to the very
compact cluster centred around the star identified as 05374631--6909100
in the 2MASS catalogue, located near the lower-left corner of
Figure\,\ref{fig1}, in chip WF\,3. In Figure\,\ref{fig3} we show the CMD
of the whole of chip WF\,3 (grey dots) compared with that of the cluster
stars (thick black dots), defined as those within three times the
apparent cluster half-light radius of $5\arcsec$ or $1.2$\,pc. Since the
cluster members are most likely coeval, the presence of considerable
colour scatter in this small region suggests that the effects of
differential extinction are significant.

As an initial attempt to estimate the range of extinction, we plot in
Figure\,\ref{fig3} the 4\,Myr isochrone of Marigo et al. (2008) used in
Figure\,\ref{fig2} and apply to it a reddening correction according to
the Galactic extinction law of Seaton (1979) for different values of
$E(B-V)$. While there is no guarantee that the average Galactic
extinction law applies in this region (in fact it does not, as we will
show in Section\,5), our purpose here is to show that CMDs are severely
dominated by variable extinction. The solid, long-dashed and
short-dashed lines correspond  respectively to colour excess values of
$E(B-V)=0, 0.3$ and $0.7$. The largest majority of MS stars seem to lie
in the colour range defined by $0.3 < E(B-V) < 0.7$, with very few of
them bluer than the $E(B-V)=0.3$ limit. The long-dashed line, defining
the blue envelope to the observed MS, can be considered as
representative of the cluster's zero-age MS. From it, we infer an
approximate initial mass for the brightest stars at $m_{606}=13.8$ of
$\sim 40$\,\Msolar. Unfortunately, a more precise mass (and age)
determination would require an accurate knowledge of the reddening
towards each star, which is not possible due to the lack of reference
evolutionary features on the MS. 

However, of particular interest for understanding the extent of patchy
absorption and the ensuing differential reddening is the inspection of
the elongated stellar sequence running almost parallel to the MS at
$m_{606} \simeq 20$ in Figure\,2 (easier to see in Panels\,b and c).
This apparent sequence is in reality the clump of red giant stars
(hereafter ``red clump'', RC), characteristic of low-mass, metal-rich
stars experiencing their He-burning phase and represents the counterpart
at high metallicity and younger ages of the horizontal branch seen in
CMDs of globular clusters. The RC usually appears as a very tight, well
defined feature in the CMD of stellar populations. For instance, in the
solar neighbourhood, {\em Hipparcos} data show the RC to extend in colour over
the range $0.8 < (V-I)_0 < 1.25$, with a mean absolute magnitude
$M_I=-0.23 \pm 0.03$ and a dispersion in magnitude of
$\sigma_\mathrm{RC}=0.20$\,mag (Stanek \& Garnavich 1998). While this
has suggested that the RC could in principle be used as a distance
indicator (Paczynski \& Stanek 1998), later theoretical works (e.g. Cole
1998; Girardi et al. 1998; Girardi \& Salaris 2001; Salaris \& Girardi
2002) have shown that this requires the knowledge of the properties of
the population being studied, since metallicity and age affect the
magnitude and colour, as well as the shape of the RC in the CMD. 

Obviously, also the intervening absorption affects the location of the 
RC in the CMD. Recently, Nataf et al. (2013) have compared differences
in the colour and apparent magnitude of RC stars towards several sight
lines in the Galactic bulge to derive the extinction law in this part of
the Galaxy. Extinction can even modify the apparent shape of the RC in
the CMD if the absorption is patchy like in our case
(Figure\,\ref{fig1}), since the amount of extinction varies from star to
star and the RC becomes in practice an elongated feature (hereafter
called ``strip'') running parallel to the direction of the reddening
vector. Fortunately, as we explain below, models show that age,
metallicity, distance and reddening act differently on the location and
the shape of the RC in the CMD. Therefore, with high precision and high
resolution photometry of a large number of stars in four or more bands
it is possible to disentangle the four effects and determine or
constrain their values. In the sections that follow we will make use of
this information to derive the extinction law in this specific region of
the LMC and to measure absolute reddening towards individual stars. 

{We start by illustrating the method, which hinges on the central
assumption that a single extinction law applies throughout this field.
Note, however, that this is not just a mere assumption but rather an
inference, since the existence of a well-defined reddening vector in
each of the CMDs (see Figure\ref{fig2}) argues for one and the same
extinction law for all the stars. }

\vspace*{-0.5cm}
\section{The red giant clump}

Girardi \& Salaris (2001) and Salaris \& Girardi (2002) have studied the
behaviour of the mean RC as a function of age and metallicity. In
particular, they have derived the average absolute magnitude and colour
in the $V$ and $I$ bands from synthetic CMDs for different combinations
of age and metallicity, as well as for different star formation
histories. In their models, stars belonging to the RC are all those that
have passed the He flash and are experiencing their central He burning
phase (Cannon 1970). Using theoretical CMDs of the same type, computed
for the specific WFPC\,2 bands of our observations, we have studied the
shape of the RC.

The theoretical CMDs are based on the simple case of a single burst of
star formation and a Salpeter initial mass function, starting from a set
of isochrones and corresponding luminosity functions for various ages
and metallicities. The grid of models covers ages in the range from
$1.4$\,Gyr through to $3.0$\,Gyr with a step of $0.4$\,Gyr and four
values of the metallicity, namely $Z=0.001, 0.004, 0.008$ and $0.019$, i.e.
roughly from $1/20\,Z_\odot$ to $Z_\odot$. In fact, red clump stars can
have ages up to 10\,Gyr, but as illustrated by Girardi \& Salaris (2001), 
the age distribution of red clump stars in galaxies with constant star 
formation is strongly skewed towards younger ages, due to the longer 
lifetimes of more massive RC stars and to the decreasing rate at which 
stars leave the MS at older ages. Moreover, red clump stars 
with ages between 3 and $\sim 9$ Gyr change very little their intrinsic 
positions in the CMDs. This justifies limiting our interval at ages 
up to 3\,Gyr. 

\begin{figure*}
\centering
\resizebox{\hsize}{!}{\includegraphics{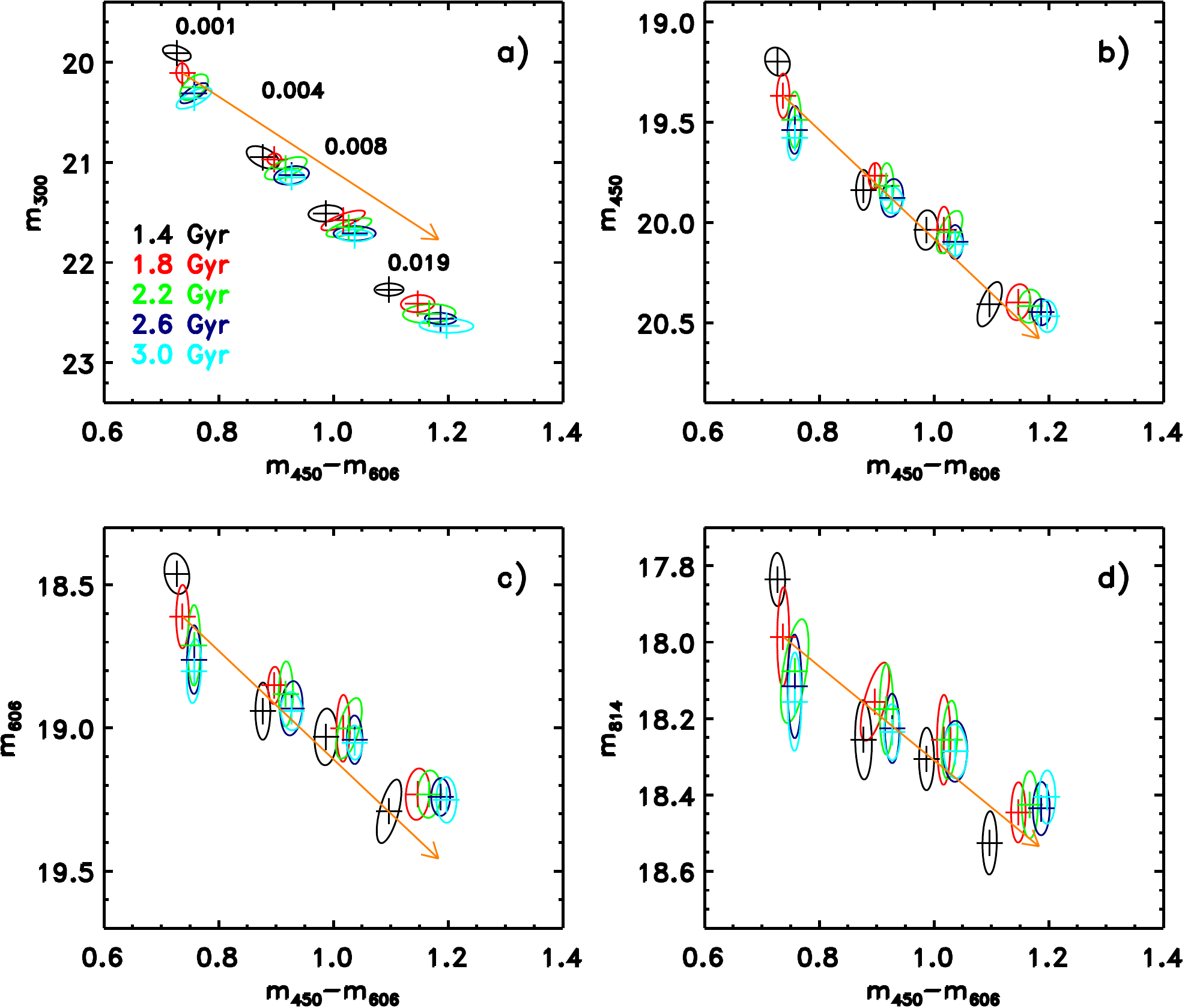}}
\caption{Red clump nominal ellipses from the models of Girardi \&
Salaris (2001) for various ages and metallicities, as indicated. The
ellipses correspond  to the 67\,\% isophotal profile measured in the
Hess diagrams. A distance modulus of $18.55$ is assumed and account is
taken of the interstellar extinction component due to the MW along the
line of sight to the LMC, corresponding to $E(B-V)=0.07$.  The shrinking
scale in the ordinates from Panel\,a) through to d) shows that the
magnitude differences progressively decrease at longer wavelengths. For
illustration purposes, the arrow shows the reddening vector
corresponding to $E(B-V)=0.4$ for the MW extinction law.}
\label{fig4}
\end{figure*}

We have characterised the location of all RC stars in the corresponding
CMDs by means of Hess diagrams, fitting a series of elliptical isophotes
to the RC in each CMDs in order to determine its parameters, namely the
colour, magnitude, semi-major axis, ellipticity and position angle. The
essence of our simulations is shown schematically in Figure\,\ref{fig4},
where we plot the theoretical RC, for each combination of age and
metallicity, in the plane of the CMDs defined by our bands. The age
ranges from $1.4$ to $3.0$\,Gyr and the metallicity from $Z=0.001$
to $0.019$, as indicated. We have selected in all panels the
$m_{450}-m_{606}$ colour, since it is similar to the commonly used
$B-V$. All ellipses shown in the figure correspond to the 67\,\%
isophotal profile and therefore define the $1\,\sigma$ uncertainty. In
order to convert absolute magnitudes to apparent values, we have 
assumed a distance modulus $(m-M)_V=18.55$, as mentioned above, as
appropriate for the neighbouring SN\,1987A field in the LMC (Panagia
1998). The resulting magnitudes have been further corrected for the
interstellar extinction component due to the MW along the line of sight
to the LMC, corresponding to $E(B-V)=0.07$ (Fitzpatrick \& Savage 1984;
see also Brunet et al. 1975 and Isserstedt 1975). The reddening corrections in
the various bands amount to $A_{300}=0.36$, $A_{450}=0.26$,
$A_{606}=0.18$, and $A_{814}=0.12$. Note that, at longer wavelengths,
the differences between the magnitude of the RC at different ages and
metallicity progressively decrease, as is witnessed by the shrinking
scale in the ordinates from panel a) through to d). 

Since both the distance and the MW foreground component of the
extinction are fixed, the ellipses of Figure\,\ref{fig4} can only move
in the CMD because of the extinction within the LMC. As an example, the
arrows plotted in Figure\,\ref{fig4} show the  reddening vector
corresponding to the MW extinction law of Seaton (1979), for a colour
excess $E(B-V)=0.4$. Figure\,\ref{fig4} is useful to understand the
combined effect of metallicity, age and extinction on the location and
shape of the RC in the CMD and can be directly compared with the
observations to derive or constrain the properties of the population.
This is particularly true when observations in several bands are
available, since Figure\,\ref{fig4} shows that metallicity, age and
reddening affect the position of RC stars in different ways in the CMDs
defined by our bands. 

For instance, age does not appear to considerably affect the magnitude
of the RC, except at very low metallicity. As for the reddening, in
bands longwards of $\sim 4\,000$\,\AA\, an increase of reddening or
metallicity move the RC along very similar directions. While the
reddening vector shown in Figure\,\ref{fig4} follows the MW extinction
law, namely $R_V=A_V/E(B-V)=3.1$, the situation would not change if the
slope of the extinction law would vary within a factor of $\pm 30$\,\%
(i.e. $2 \la R_V \la 4$). Therefore, if the absolute value of the
reddening is known and so is the distance, the position of the observed
RC in these bands can constrain the metallicity of the population
leaving some uncertainties on the age. Interestingly, this limited
sensitivity of the RC to age and metallicity makes it easier to estimate
the amount of reddening when there is considerable interstellar
extinction ($A_V>1$), since in this case the magnitude and colour
displacement of the RC in the CMD due to extinction dominates over all
uncertainties on metallicity and age. We will address this case
specifically in Section\,5. When observations at shorter wavelength are
available (e.g. Figure\,\ref{fig4}a), it also becomes possible to break
the degeneracy between reddening and metallicity. Provided that the
distance is known, both the extinction law and reddening value $A_V$
can be derived.

\begin{figure*}
\centering
\resizebox{\hsize}{!}{\includegraphics[angle=90]{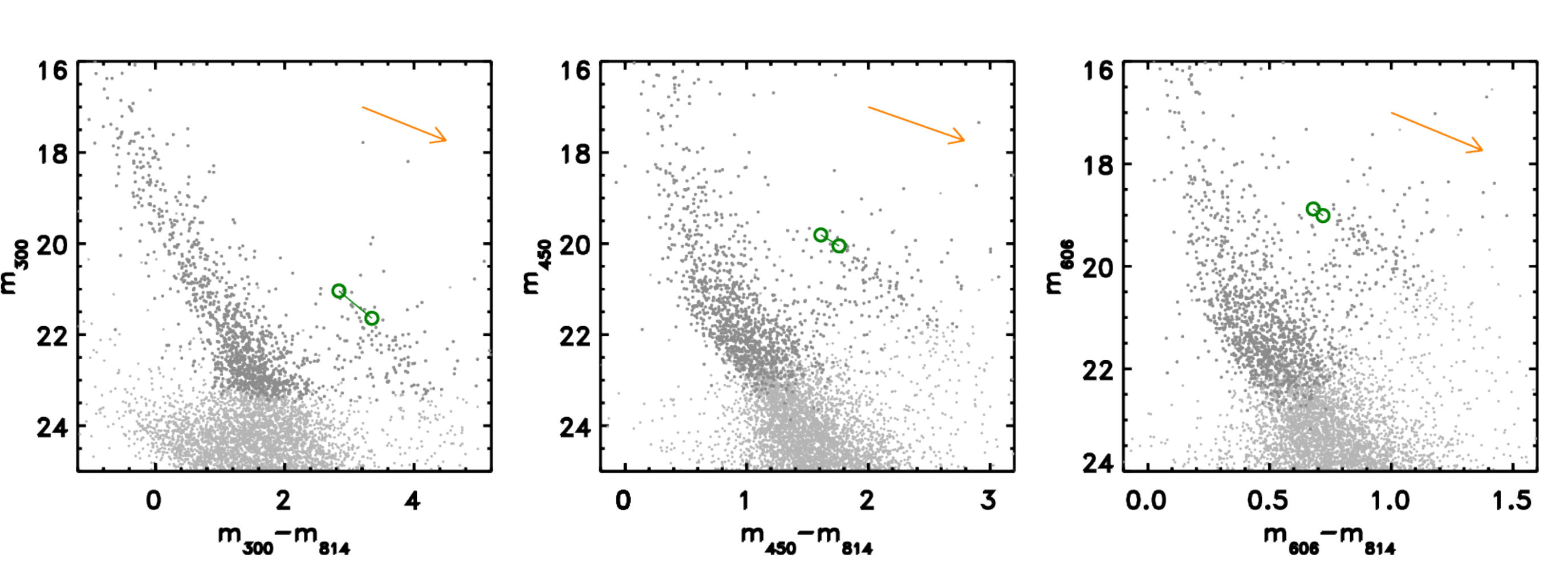}}
\caption{The CMD of the stars in our field is compared with the
theoretical positions of the RC for metallicity $Z=0.004$ and $0.008$
(left and right, respectively), having assumed a distance modulus 
$(m-M)_V=18.55$ and a MW contribution to the extinction along the line
of  sight of  $E(B-V)=0.07$. For illustration purposes, the arrows  show
the reddening vector for the MW extinction law with $E(B-V)=0.4$.}
\label{fig5}
\end{figure*}

\vspace*{-0.5cm}
\section{Measuring the extinction for red clump stars}

The theoretical location of the RC predicted by our simulations is shown
together with the observations in Figure\,\ref{fig5}, from which some
preliminary qualitative information on the old population present in
this field can be drawn. The thick green circles show the position of
the RC for an age of $2.2$\,Gyr and, from left to right, the two values 
of the metallicity defining the current range of accepted values for the
LMC, namely $Z=0.004$ and $0.008$ (e.g. Hill, Andrievsky \& Spite
1995; Geha et al. 1998). The choice of band combinations in the CMDs of
Figure\,\ref{fig5} always include the F814W band, since its photometry
is most accurate. An inspection of Panels\,b) and c), where the
photometric uncertainties are smaller, immediately reveals that age and
metallicity alone cannot explain the elongated shape of the RC, which
requires instead a large spread of reddening values to be reproduced.
For comparison, the arrow corresponds to a reddening spread of
$E(B-V)=0.4$, for the Galactic extinction law, but the extension of the
RC suggests that the actual spread is even higher.

Given this considerable range of extinction values, the CMD in Panel\,a)
suggests that the metallicity of the population is not likely to extend
beyond the $0.004 < Z < 0.008$ range, since this would require a colour
spread larger than observed for RC stars. This is consistent with the
average metallicity of field stars in the LMC being  $Z \simeq 0.007$
(e.g. Hill, Andrievsky \& Spite 1995; Geha et al. 1998), but it confirms
that large spreads may exist and that lower metallicity stars are likely
present in this region (according to Dopita et al. 1997 the LMC
metallicity might have almost doubled over the past 2\,Gyr). As regards
the age of RC stars, comparison between the observations and models in
Panel\,c) suggests that it is most likely in excess of 1\,Gyr, given the
limited magnitude spread of the RC band.

In order to proceed to a more quantitative analysis of interstellar
extinction in this field, it is necessary to determine more precisely
which stars belong to the RC and to calculate their displacement in the
CMD from the location that they would occupy if there were no
extinction. For a starting location, which we hereafter will call
``nominal RC'', we take the theoretical RC of stars of the lowest
metallicity consistent with the data, in the sense that for all ages in
our range of models the nominal RC should not be redder than the
observed RC stars, in all bands simultaneously. In the case of our
field, Figure\,\ref{fig5} suggests that the most appropriate metallicity
is $Z=0.004$. This choice defines in practice the magnitude of the
nominal RC, given the limited spread in luminosity within each
metallicity class (see Figure\,\ref{fig4}c). As for its colour, we take
it to be the average RC colour for ages in the range $1.4 - 3.0$\,Gyr,
since lower ages are less likely as Figure\,\ref{fig5}a shows.

As for the uncertainties on the nominal RC colour and magnitude, we have
taken the uncertainties at each age for $Z=0.004$ and combined them in
quadrature. The corresponding magnitudes, already including the effects
of intervening MW extinction, are $m_{300}=21.11 \pm 0.12$ and
$m_{450}=19.81 \pm 0.12$, $m_{606}=18.89 \pm 0.09$,  $m_{814}=18.21 \pm
0.10$, and the colours in the most common band combinations are
$m_{300}-m_{450}=1.3 \pm 0.1$, $m_{450}-m_{606}=0.92 \pm 0.06$ and
$m_{606}-m_{814}=0.68 \pm 0.04$, with all uncertainties given at the
$1\,\sigma$ level. The resulting nominal RC, defined in this way, is
shown as an example by the ellipse in Figure\,\ref{fig6} in the
$m_{606}$ {\em vs.} $m_{606}-m_{814}$ CMD, where photometric
uncertainties are smallest. The semi-major and semi-minor axes of the
ellipse are, respectively, $2.5$ times the uncertainty on the RC
magnitude and colour. 

\vspace*{-0.4 cm}
\subsection{Identifying red clump stars}

\begin{figure}
\centering
\resizebox{\hsize}{!}{\includegraphics{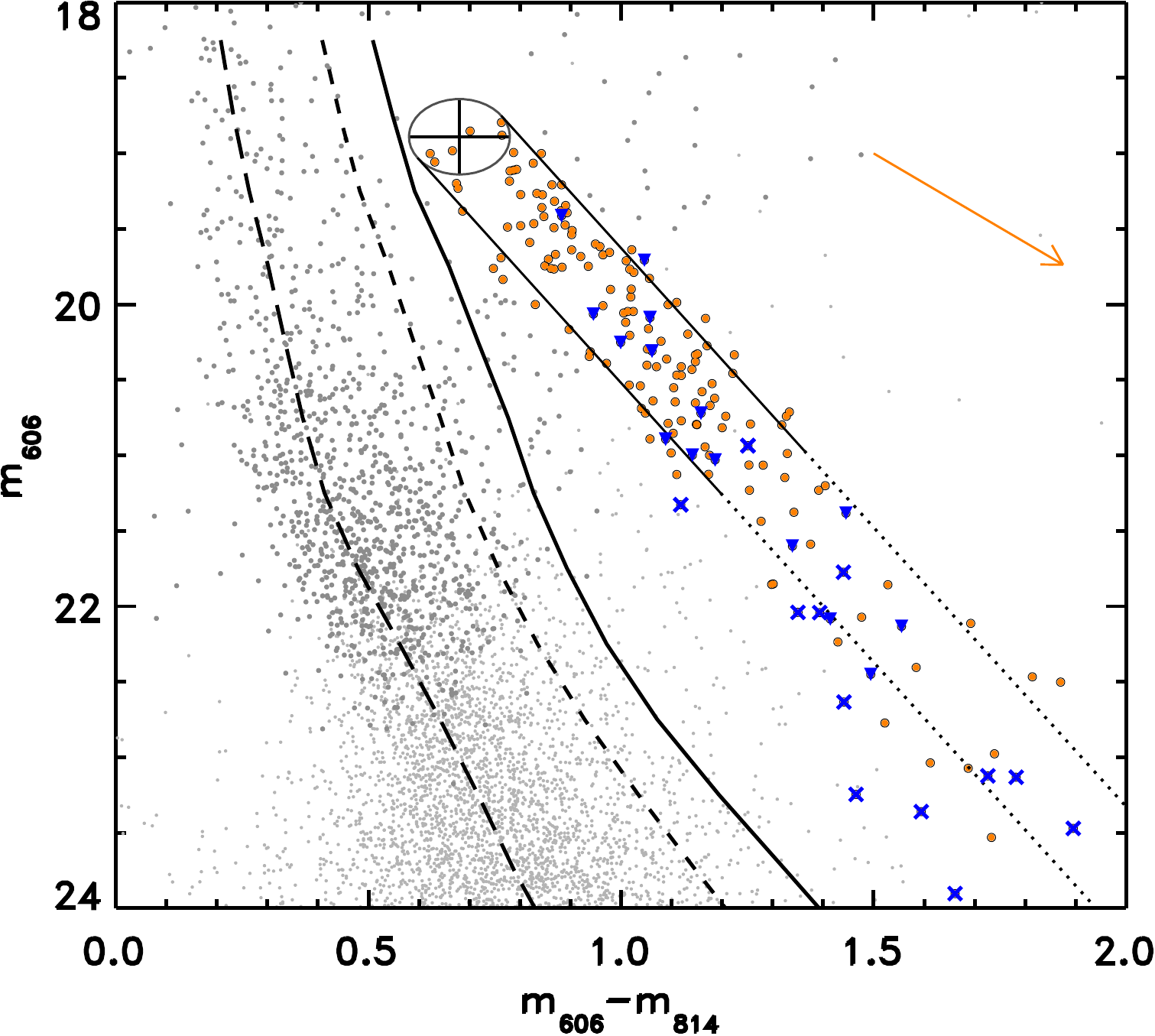}}
\caption{The nominal position of the RC, already including the effects
of MW reddening, is shown by the ellipse. To select reddened RC stars we
have excluded all objects bluer than the solid line, which are closer to
the MS ridge line (dashed line) than three times the width of its colour
distribution ($3\,\sigma$). Thick orange dots 
inside the region marked by the solid straight lines are candidate RC
stars, whereas objects indicated with crosses or triangles are likely
PMS objects, which we have excluded from our sample. As in
Figures\,\ref{fig4} and \ref{fig5}, the  arrow shows the reddening
vector for the MW extinction law, with $E(B-V)=0.4$.}
\label{fig6}
\end{figure}

The first condition to be met by true RC stars in the CMDs of
Figure\,\ref{fig6} is that they must be redder and fainter than the
nominal RC ellipse. We begin by considering as candidate RC stars all
objects redder than the MS by at least three times its colour width. The
dashed line shows the MS ridge line, determined by iteratively removing
from each magnitude bin all stars with a colour departing by more than
$2.5\,\sigma$ from the median. The solid curve corresponds to the
$3\,\sigma$ envelope and we consider as candidate RC stars all objects
redder than this curve.

The effect of differential extinction is to spread the stars contained
in the nominal RC ellipse throughout the CMD, according to the
extinction law, thus forming a strip of higher stellar density. Also
red giant stars brighter and fainter than those in the RC are displaced
by reddening in the CMD plane and must be excluded from our analysis.
However, if the extinction law is uniform (i.e. the same or very similar
$A_\lambda/E(B-V)$ ratio applies to all objects in the field), only 
limited flaring is expected along the RC strip, which remains separate
from other RG stars. This is easiest to see in the $m_{606}$ {\em vs.}
$m_{606}-m_{814}$ CMD of Figure\,\ref{fig6}, because photometric errors
are small and, as Figure\,\ref{fig4} suggests, the effect of reddening
dominates. 

As a first attempt to isolate bona fide RC stars, we trace two parallel
lines, tangent to the ellipse and flanking the RC ``beam'' (shown by the
lines tangent to the ellipse in Figure\,\ref{fig6}) and count the number
of stars contained within the area delimited by these lines, by the
nominal RC ellipse at the left-hand side and by the condition
$m_{606}-m_{814} \le 2$ at the right-hand side, as a function of the
lines slope. Stars falling outside of the beam by more than $2.5$ times
their colour and magnitude uncertainties are excluded. All stars
indicated by thicker marks (dots, triangles and crosses) are candidate
RC objects, although as we will show later we will limit the selection
to stars corresponding to the dots brighter than $V \simeq 23.5$ (inside
the region defined by the solid straight lines), to include only stars
with photometric uncertainty on the $m_{300}$ magnitude of less than
$0.4$\,mag and excluding any objects indicated with triangles or crosses
as they might be pre-main sequence (PMS) stars.

\begin{figure}
\centering
\resizebox{\hsize}{!}{\includegraphics{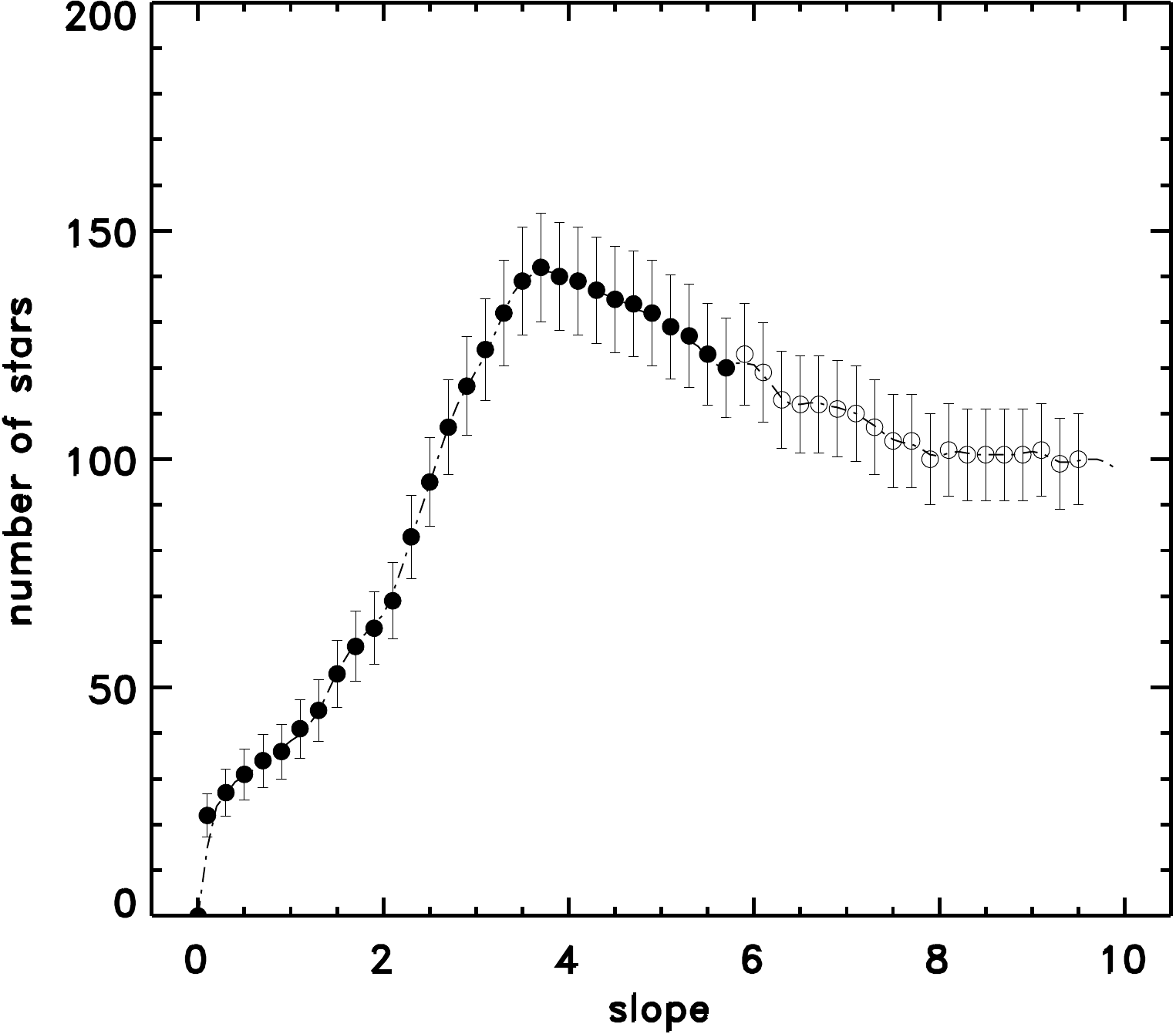}}
\caption{Number of stars contained within the RC ``beam'' as a function
of the beam's slope. This is used to show the slopes corresponding to
the peak and minimum of the distribution. This is only done for the
CMD in the F606W and F814W bands.}
\label{fig7}
\end{figure}

The number of stars inside the beam is shown in Figure\,\ref{fig7} as a
function of the slope. The slope that maximises the number of stars
within the beam (used to draw the lines in Figure\,\ref{fig6}) gives a
first indication of the extinction law in the form of the ratio
$A_{606}/E(m_{606}- m_{814})$. With this choice of the slope, we find a
total of 170 over the entire  magnitude range, which we consider as
candidate RC stars. All stars indicated by thicker marks in
Figure\,\ref{fig6} (i.e. dots, triangles and crosses) are candidate RC
objects, although as we will show later we will limit the selection to
stars corresponding to the dots brighter than $V \simeq 23.5$, since
objects indicated with triangles and crosses might be PMS stars.

After this initial selection of RC candidates, the procedure to arrive
at the final selection of bona fide RC stars, and at the determination
of the reddening slopes in the CMDs, is as follows. 

\begin{figure*}
\centering
\resizebox{\hsize}{!}{\includegraphics[angle=90]{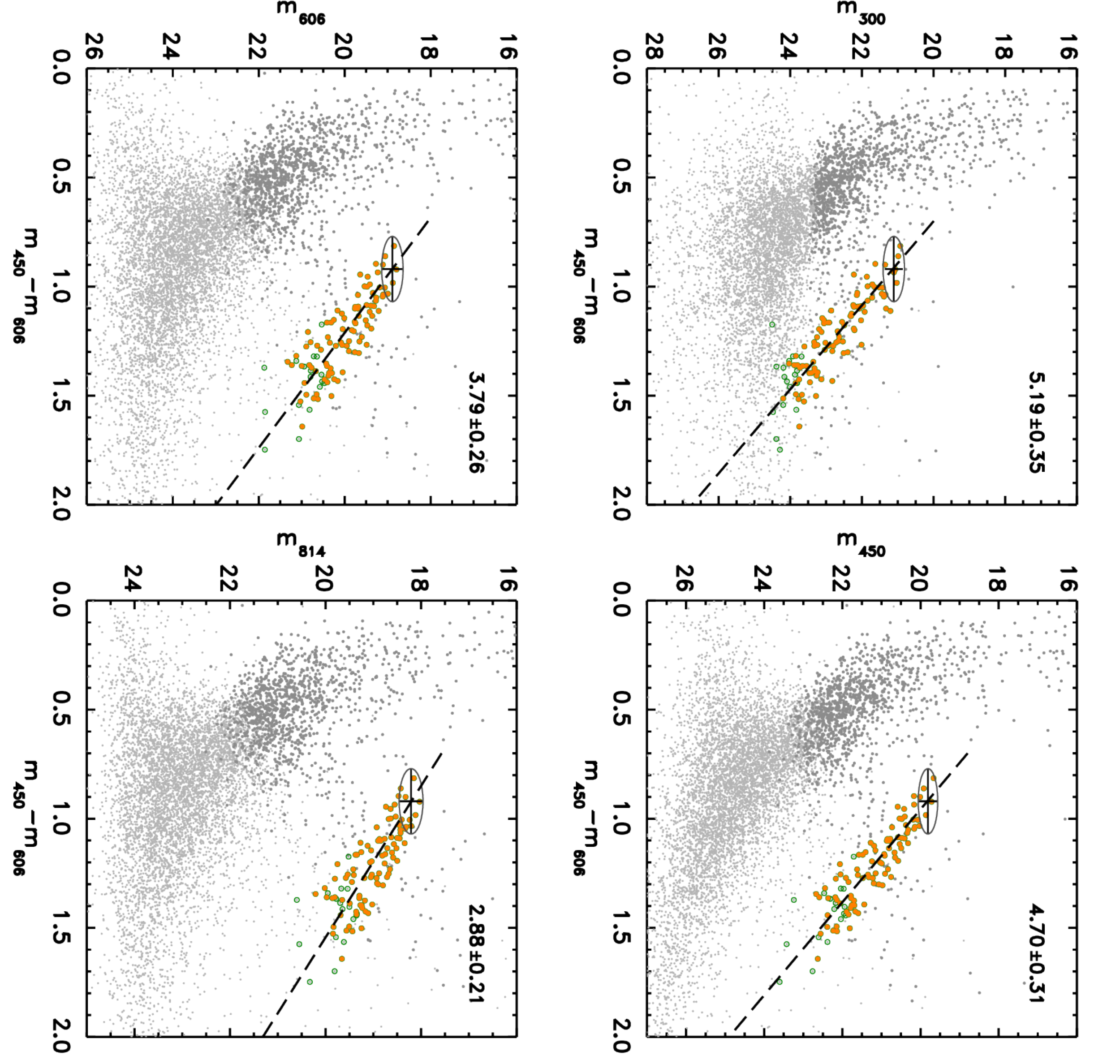}}
\caption{Bona fide RC stars. The thick orange dots are the the 90 RC stars selected with a stricter $\delta_4 <
0.15$ condition, whereas the green circles 
are those with $0.15 < \delta_4 < 0.23$. The best fitting slopes and
their uncertainties indicated in each panel refer to the 90 RC objects
with stricter condition $\delta_4 < 0.15$.}
\label{fig8}
\end{figure*}

\begin{enumerate}

\item
We consider all CMDs in which the magnitudes of the stars are  plotted
as a function of the $m_{450}-m_{606}$ and $m_{606}-m_{814}$ colours, for a
total of 8 different CMDs. An example is shown in Figure\,\ref{fig8}. 

\item 
In each CMD we derive the best linear fit to the distribution of the 170
candidate objects, taking into account: {\em 1)} the uncertainties on
their  magnitudes and colours; {\em 2)} the uncertainties on the
ellipse defining the model RC in each individual CMD; and {\em 3)} an
equal number ($=170$) of synthetic unreddened RC stars with a Gaussian
distribution inside the ellipses. The latter step guarantees that the 
best linear fit passes through the centre of the nominal RC. We derive
in this way the slope $q$ of the best fit in each CMD, and the
corresponding uncertainties $\sigma_q$.

\item
In each CMD, we identify all objects inside the region defined by the
nominal RC ellipse and two lines tangent to it having slopes corresponding
to $q \pm \sigma_q$, and consider all stars with $\delta_4 < 0.15$
inside it (or $\delta_4 < 0.23$ for a coarser selection). We include
also objects formally outside of this region but whose photometric error
ellipses in the CMD would in part overlap with it. 

\item
The selection of objects identified in the previous step is further
reduced by considering only objects whose $m_{300}$ magnitude uncertainty 
is less than $0.4$\,mag. This reduces the sample to 107 bona-fide RC 
stars with $\delta_4 < 0.23$ (or 90 bona-fide RC objects with $\delta_4 
< 0.15)$. 

\item  Finally, the procedure in step (ii) above is repeated on the
bona-fide RC stars selected in this way in order to derive the slopes
and uncertainties of the best fits in all bands, as shown for instance
in Figure\,\ref{fig7}.

\end{enumerate}

The reddening slopes derived in this way correspond to the value of the
ratio $R$ between absolute and selective extinction in the specific bands
of our observations. The values of the ratio $R$ and the corresponding
uncertainties as are listed in Table\,\ref{tab2}. 

\begin{table}
\centering 
\caption{Measured values of the ratio $R$ between absolute ($A$) and 
selective ($E$) extinction in the specific bands of our observations, 
with corresponding uncertainties. The effective wavelength ($\lambda$)
and wavenumber ($1/\lambda$) of each band are also indicated.}
\begin{tabular}{cccc} 
\hline
Band combination & $R$ & $\lambda$ [\AA] & $\mu$m$^{-1}$\\
\hline
$A_{300}/E(m_{450}-m_{606})$ & $5.19 \pm 0.35$ & 2\,979 & $3.36$ \\
$A_{300}/E(m_{606}-m_{814})$ & $5.51 \pm 0.28$ & 2\,979 & $3.36$ \\
$A_{450}/E(m_{450}-m_{606})$ & $4.70 \pm 0.31$ & 4\,531 & $2.21$ \\
$A_{450}/E(m_{606}-m_{814})$ & $5.05 \pm 0.35$ & 4\,531 & $2.21$ \\
$A_{606}/E(m_{450}-m_{606})$ & $3.79 \pm 0.26$ & 5\,938 & $1.68$ \\
$A_{606}/E(m_{606}-m_{814})$ & $4.03 \pm 0.29$ & 5\,938 & $1.68$ \\
$A_{814}/E(m_{450}-m_{606})$ & $2.88 \pm 0.21$ & 7\,942 & $1.26$ \\
$A_{814}/E(m_{606}-m_{814})$ & $3.11 \pm 0.24$ & 7\,942 & $1.26$ \\
\hline      
\end{tabular}
\vspace{0.5cm}
\label{tab2}
\end{table} 

\subsection{Possible interlopers}

An important step in assembling a reliable sample of bona-fide RC stars
is the exclusion of interlopers. The latter are objects whose colours and
magnitudes would place them inside the RC strip without their necessarily
being sources of this type. An obvious class of possible contaminants is
PMS stars, which in light of their younger age are
systematically redder and brighter than MS objects. In their study of
PMS stars in 30 Dor, De Marchi et al. (2011b) specifically excluded the
region of the CMD where PMS and RC objects would overlap. In order to
keep our sample of RC stars as much as possible free from PMS objects we
have looked for stars with H$\alpha$ excess emission, which is a
characteristic signature of accreting PMS stars. 

Employing the method developed by De Marchi et al. (2010; 2011a) in
order to identify stars with H$\alpha$ excess emission, we have compared
the $V-H\alpha$ colours of all stars in the RC strip with those of all
other stars in the field, as a function of their $V-I$ colours. The
majority of stars in the field do not have H$\alpha$ excess emission. By
selecting those with smallest combined photometric uncertainty in the
$V$, $I$ and $H\alpha$ bands ($\delta_3 < 0.05$, see
Equation\,\ref{eq6}), we have derived the reference template providing
the $V-H\alpha$ colour of non-emitting stars as a function of their
$V-I$ colours. Any excess in $V-H\alpha$ with respect to the reference
template can be converted into the corresponding equivalent  width
$W_{\rm eq}(H\alpha)$ using the relationships provided in De Marchi et
al. (2010). 

Chromospheric activity is a likely source of H$\alpha$ emission in
low-mass stars, due to non-radiative heating mechanisms powered by
magnetic fields. In the outer layers of these stars the temperature
increases towards the surface and the main cooling mechanism is
radiative loss through strong resonance lines, such as indeed H$\alpha$
(e.g. Linsky 1980). Therefore, it is not surprising if some of the RC
stars in our sample show H$\alpha$ excess emission from chromospheric
activity. However, for objects of spectral type earlier than M the value
of  $W_{\rm eq}(H\alpha)$ vis not expected to exceed a few \AA\ and if
larger values are measured mass accretion must be present. Martin (1998)
and more recently White \& Basri (2003) have proposed thresholds above
which stars with H$\alpha$ excess emission should be considered
classical T Tauri objects. More precisely, as regard the spectral range
applicable to our RC sources, White \& Basri (2003) propose thresholds
of 3\,\AA\ for stars of spectral type K\,0 -- K\,5 and values larger
than 10\,\AA\ for K\,7 -- M\,2.5 stars. 

We cannot assign an accurate spectral type or effective temperature to
our objects, since we do not know their individual reddening values, but
we can assume a typical reddening. We take $E(B-V) \simeq 0.2$. as this
is the value given by Panagia et al. (1987) and Scuderi et al. (1996)
for the region around SN\,1987A and which, as we will see in Section\,6,
is also the average value for our region. In this case, almost all stars
in the RC strip would have spectral type earlier than M\,2.5, as they
would have $(V-I)_0 < 1.5$, and those with $(V-I)_0 < 1$ would have
spectral type K\,5 or earlier. In Figure\,\ref{fig6} we have indicated
with crosses all stars with  10\,\AA $< W_{\rm eq}(H\alpha) < 30$\,\AA\ 
and with triangles those with 3\,\AA $< W_{\rm eq}(H\alpha) < 10$\,\AA. 

The former must be excluded from our sample as they are likely PMS stars
and do not belong to the bona-fide RC population. Interestingly, they
represent about 1/3 of the candidate objects in the RC strip with $V-I
\ga 1.3$, in the portion of the strip defined by the dotted lines in
Figure\,\ref{fig6}. In fact, all these objects have already been
excluded from our sample since they all have uncertainties in the
$m_{300}$ magnitudes exceeding $0.4$\,mag, as discussed above. However,
as mentioned above  there are also two objects with $W_{\rm eq}(H\alpha)
> 10$\,\AA\ in the sample of 107 bona-fide RC stars with $\delta_4 <
0.23$ (respectively with $W_{\rm eq}(H\alpha)$ of 14\,\AA\, and
27\,\AA), which we will remove as they too are likely PMS stars not
related with the RC sample. 

As for the stars with 3\,\AA $< W_{\rm eq}(H\alpha) < 10$\,\AA\
(triangles), we have removed 10 of them from the same sample, since
some might be PMS stars. However, these objects represent less than
10\,\% of the total population and, more importantly, their placement
inside the RC strip does not reveal any peculiarity in colour or
magnitude distribution, compared with those of the objects with no
H$\alpha$ excess. This means that their presence would anyhow not affect
the determination of the slope of the reddening vector that we derive in
the following section.

\begin{figure*}
\centering
\resizebox{\hsize}{!}{\includegraphics{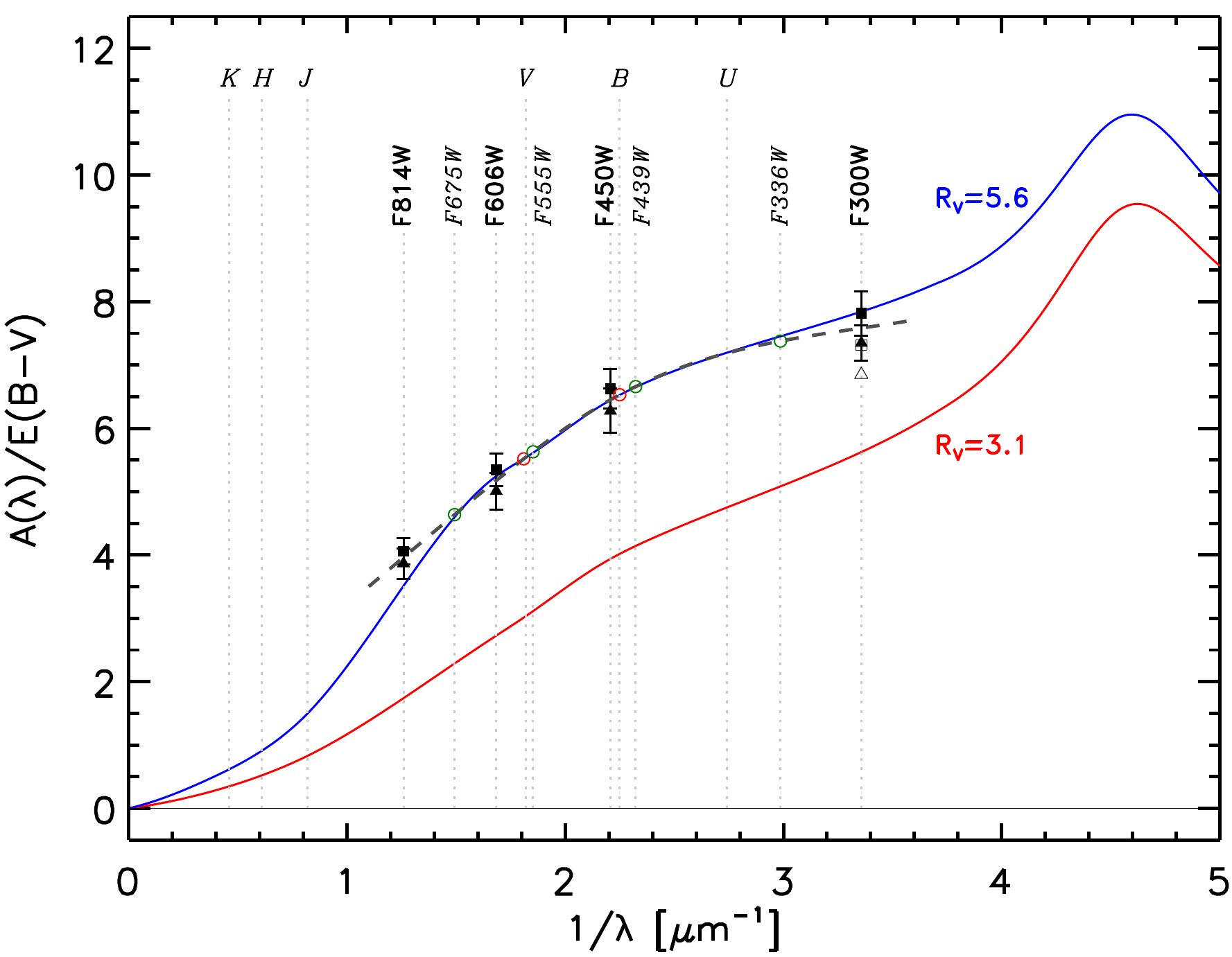}}
\caption{Extinction law. The dashed line is a spline interpolations
through our measurements, with the filled squares representing those 
derived from the CMDs as a function of $m_{450}-m_{606}$ and filled 
triangles for the CMDs as a function of $m_{606}-m_{814}$. The circles 
are the interpolated values for the other WFPC\,2 bands and for the 
Johnson $B$ and $V$ filters. The extinction law is in good agreement,
within the uncertainties, with the parametric law of Fitzpatrick \&
Massa (1999) for $R_V=5.6$ and is considerably shallower than the
canonical Galactic extinction law ($R_V=3.1$).}
\label{fig9}
\end{figure*}

\vspace*{-0.3cm}

\subsection{Extinction law}

As mentioned in the Introduction, it is customary to express the
extinction law in the form of the ratio

\begin{equation}
R_\lambda \equiv \frac{A_\lambda}{E(B-V)},
\label{eq7}
\end{equation}

\noindent  
where $A_\lambda$ is the extinction at the specific wavelength or band
considered and $E(B-V)$ the colour excess in Johnson's $B$ and $V$
bands. The $R$ values listed in Table\,\ref{tab2} are expressed as a
function of the colour excess in the specific WFPC\,2 bands used in our
observations. However, spline interpolation through the values listed in
Table\,\ref{tab2} allows us to derive $A_B$ and $A_V$ as a function of
$E(m_{450}-m_{606})$ and $E(m_{606}-m_{814})$. We find
$A_B/E(m_{450}-m_{606})=4.75$, $A_V/E(m_{450}-m_{606})=4.04$, 
$A_B/E(m_{606}-m_{814})=4.30$, and $A_V/E(m_{606}-m_{814})=5.11$. Taking
into account that Equation\,\ref{eq7} implies that $R_B - R_V =1$, we
can translate the values of Table\,\ref{tab2} into the corresponding
values as a function of $E(B-V)$. The values of $R_\lambda$ found in
this way are listed in Table\,\ref{tab3} and shown graphically in
Figure\,\ref{fig9}, where filled squares and filled triangles correspond
to values obtained from measurements in the planes as a function of 
$m_{450}-m_{606}$ and $m_{606}-m_{814}$ colours, respectively. Within
the uncertainties, the two sets of values are in excellent agreement
with one another. 

Note that for the F300W band it is necessary to take into account the
considerable red-leak affecting magnitudes measured in this band (Lim et
al. 2009). Reddened stars appear less attenuated than they would be
without red-leak because the filter allows some of the flux at longer
wavelengths ($6\,000$ \, -- \, $10\,000$\,\AA) to pass. For a given
star, the magnitude difference between  F300W and an otherwise identical
filter with no transmission above  5\,000\,\AA\ depends on the
extinction law and on the effective temperature of the star, being
stronger for a shallower extinction curve (larger $R_V$ value) and for
cooler stars. To quantify this effect, we have used the {\em HST} synthetic
photometry simulator {\em Synphot} (Laidler et al. 2005) to compare the
expected magnitude of a red giant star of $T_{\rm eff}=5\,250$\,K in the
F300W band and in a band identical to it but with no transmission above
5\,000\,\AA. The magnitudes were calculated for different values of
$E(B-V)$, having assumed the extinction curve of Fitzpatrick \& Massa
(1999; in turn based on the original parametrization of Cardelli et al.
1989) for a value of $R_V=5.6$, since this is the value suggested for
this field by the other bands in our study (see below). We found in this
way that the value of $A_{300}/E(B-V) \simeq 7.1$ that we measure (empty
square and triangle in Figure\,\ref{fig9}, see also Table\,\ref{tab3})
would in fact be $\sim 7.6$ if the F300W band had no red-leak.

\begin{table}
\centering 
\caption{Measured and interpolated ({\em italics}) values of
$R_\lambda$  for various bands. The effective wavelengths ($\lambda$)
and wave numbers ($1/\lambda$) of each band are also indicated. Note
that all interpolated values, shown in italics, are given for the
specific  monochromatic effective wavelength indicated without
considering the width of the filter. The $R_\lambda$ value in the F300W
band is corrected for the effects of red leak, as explained in the text.}
\begin{tabular}{cccc} 
\hline
Band & $R_\lambda$ & $\lambda$ [\AA] & $1/\lambda$ [$\mu$m$^{-1}$]\\
\hline
F300W       & $7.58 \pm 0.33$ & ~2\,979 & $3.36$ \\
{\em F336W} & $\mathit{7.38}$ & ~3\,349 & $2.99$ \\
$U$         & $\mathit{7.19}$ & ~3\,650 & $2.74$ \\
{\em F439W} & $\mathit{6.66}$ & ~4\,305 & $2.32$ \\ 
$B$         & $\mathit{6.52}$ & ~4\,450 & $2.25$ \\
F450W       & $6.45 \pm 0.25$ & ~4\,531 & $2.21$ \\
{\em F555W} & $\mathit{5.63}$ & ~5\,398 & $1.86$ \\
$V$         & $\mathit{5.53}$ & ~5\,510 & $1.82$ \\ 
F606W       & $5.18 \pm 0.24$ & ~5\,938 & $1.68$ \\ 
{\em F675W} & $\mathit{4.64}$ & ~6\,697 & $1.49$ \\
F814W       & $3.96 \pm 0.14$ & ~7\,942 & $1.26$ \\
$J$         & $\mathit{1.50}$ & 12\,200 & $0.82$ \\ 
$H$         & $\mathit{0.92}$ & 16\,300 & $0.61$ \\ 
$K$         & $\mathit{0.61}$ & 21\,900 & $0.46$ \\ 
\hline      
\end{tabular}
\vspace{0.5cm}
\label{tab3}
\end{table}

The dashed line in Figure\,\ref{fig9} is a spline interpolation through
the $R_\lambda$ values obtained from our analysis (see
Table\,\ref{tab3}) and it agrees, within the uncertainties, with the
extinction law of Fitzpatrick \& Massa (1999) for a value of $R_V=5.6$.
In Table\,\ref{tab3} we give the values of $R_\lambda$ at the effective
wavelengths of a number of {\em HST} and standard photometric bands. The
wavelengths are also shown graphically by the vertical dotted lines in
Figure\,\ref{fig9}.

With $R_V=5.6$, the extinction law for this field is considerably less
steep than the canonical Galactic extinction law, with $R_V=3.1$, which
is also shown in Figure\,\ref{fig9}. The $R_V$ value that we find is also
considerably less steep than those reported by Gordon et al. (2003) in
the LMC. These authors combined observations with the {\em International
Ultraviolet Explorer} (discussed in Fitzpatrick \& Massa 1990), and
photometry from ground-based facilities at optical and infrared
wavelengths, to derive the extinction law towards 19 sight lines in the
LMC, from UV to near-infrared wavelengths. Note that, in
order to determine the absolute extinction from the measured colour
excess, Gordon et al. (2003) adopted the customary approximation
$R_V=1.1 \times E(V-K)/E(B-V)$, which is reliably valid only for $R_V
\simeq 3$ (Bouchet et al. 1985; Fitzpatrick \& Massa 2007). The average
$R_V$ value that they find for eight stars in the LMC\,2 supershell near
the 30 Dor star formation region is $2.76 \pm 0.09$, while the average
over ten other LMC lines of sight gives a higher value, $R_V = 3.41 \pm
0.06$. 

Our larger $R_V$ value corresponds to a ``flatter'' extinction law,
which in turn implies that the dust grains size distribution is skewed
towards larger grains, compared with the canonical {average Galactic
extinction law in the diffuse interstellar medium} (e.g. Cardelli et al.
1989; Kim, Martin \& Hendry 1994). Put differently, this implies that
larger grains dominate {(e.g. van de Hulst 1957)}. {At optical
wavelengths the  extinction curve that we find follows closely the
Galactic curve, with a grey offset of about $2.5$. This offset indicates
the presence of grains relatively large, with sizes roughly of the order
of $\lambda / (2 \, \pi)$.} A large $R_V$ value appears to be typical of
denser environments (e.g. Johnson \& Borgman 1963; Cardelli et al. 1989;
Weingartner \& Draine 2001 and references therein) and this could
suggest that a denser environment favours either a more efficient
formation of larger grains or the survival of larger grains to the
detriment of the smaller ones.

Thus, it is possible that this  region is simply denser than those
probed by the individual lines of sight sampled by Gordon et al. (2003)
and, therefore, the $R_V=5.6$ value that we find might be characteristic
of just this region. More in general, it could be the result of
systematically different conditions of the environment in the LMC,
compared to the MW. We will address this issue in a series of
forthcoming papers (De Marchi \& Panagia in preparation) devoted to the
reddening distribution in the centre of the 30\,Dor region, as well as
in the area surrounding it. Recent high-quality {\em HST} observations exist
in for these regions at UV, optical and near-infrared
wavelengths and they reveal a prominent extended RC in the CMDs (De
Marchi et al. 2011b; Sabbi et al. 2013). These observations cover also
the $J$ and $H$ bands, which will allow us to measure the properties of
the extinction law in the near infra-red and to further validate the
$R_V=5.6$ value obtained in this work.

\vspace*{-0.2cm}
\section{Reddening distribution}

Having derived the extinction law applicable to the stars in this field,
it is possible to measure the extinction towards individual objects when
the nominal location of these stars in the CMD can be accurately 
determined, like in the case of RC stars. We will consider separately RC
objects and stars in the UMS, in order to study the relative
distribution of stars and dust in this field. Using data from the
Magellanic Clouds Photometric Survey (Harris, Zaritsky \& Thompson
1997), Zaritsky (1999) concluded that there is a significant difference
in the extinction towards hot ($T_{\rm eff} > 12\,000$\,K) and cold
(5\,500\,K $< T_{\rm eff} <$ 6\,500\,K) stars, with the hotter stars being
more highly extincted on average. As we will show, this conclusion does
not apply to the objects in this specific field.

\begin{figure}
\centering
\resizebox{\hsize}{!}{\includegraphics{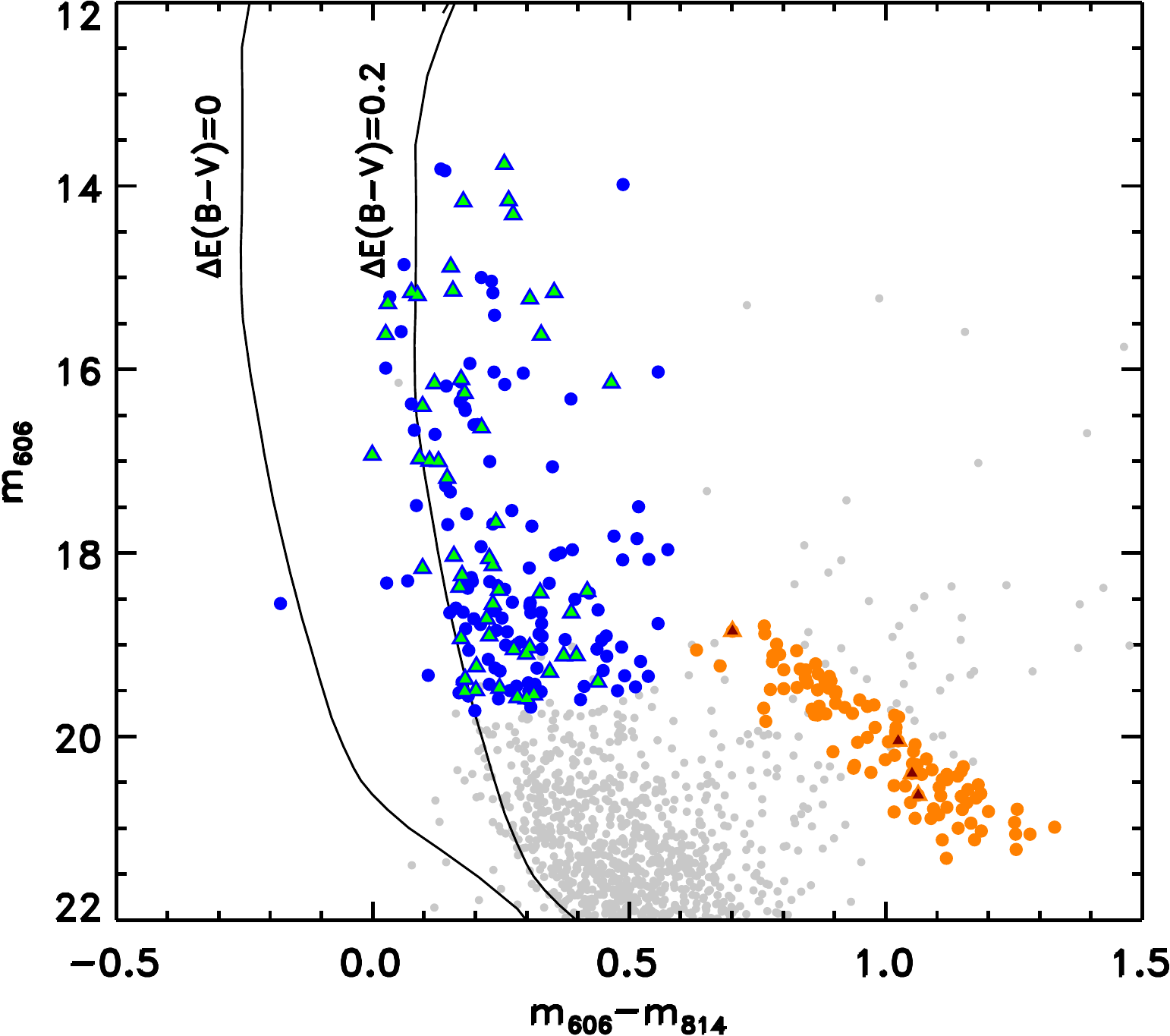}}
\caption{The stars indicated by thick dots are used to study the
distribution of reddening in the field. Triangles correspond to stars 
located inside a radius of $15\arcsec$ ($3.75$\,pc) of the small young
cluster in the field. The left-hand curve is the 4\,Myr old isochrone with
only MW extinction, namely $E(B-V)=0.07$, applied  to it, while the
solid curve on the right also includes an additional $\Delta
E(B-V)=0.2$  applied according to the extinction law derived in this
work.}
\label{fig10}
\end{figure}

The objects that we consider in this study are indicated as thick
symbols in Figure\,\ref{fig10}. Besides the bona-fide RC stars discussed
in the previous section, we have also considered all UMS stars that are
brighter than the 20$^{\rm th}$ magnitude in all four bands. Objects
indicated by triangles are those within a radius of $15\,\arcsec$
($3.75$\,pc) from the centre of the small young cluster in the field
(see Figure\,\ref{fig1}). They represent 1/3 of the UMS objects
in the field, although four RC stars are projected in the same
area.  Also shown in Figure\,\ref{fig10} is the isochrone for a
population with metallicity $Z=0.008$ and age of 4\,Myr from the models
of Marigo et al. (2008). The isochrones are shown for two values of the
reddening: the one on the left only includes the intervening MW
absorption along the line of sight to the LMC, namely $E(B-V)=0.07$ as
mentioned above, without any additional LMC reddening, while the one on
the right has an additional reddening $\Delta E(B-V)=0.2$, applied
according to the extinction law derived in the previous
section.\footnote{Hereafter, we indicate with $\Delta E(B-V)$ the
reddening in addition to the contribution of the MW along the line of
sight to the LMC.} 

\begin{figure}
\centering
\resizebox{\hsize}{!}{\includegraphics{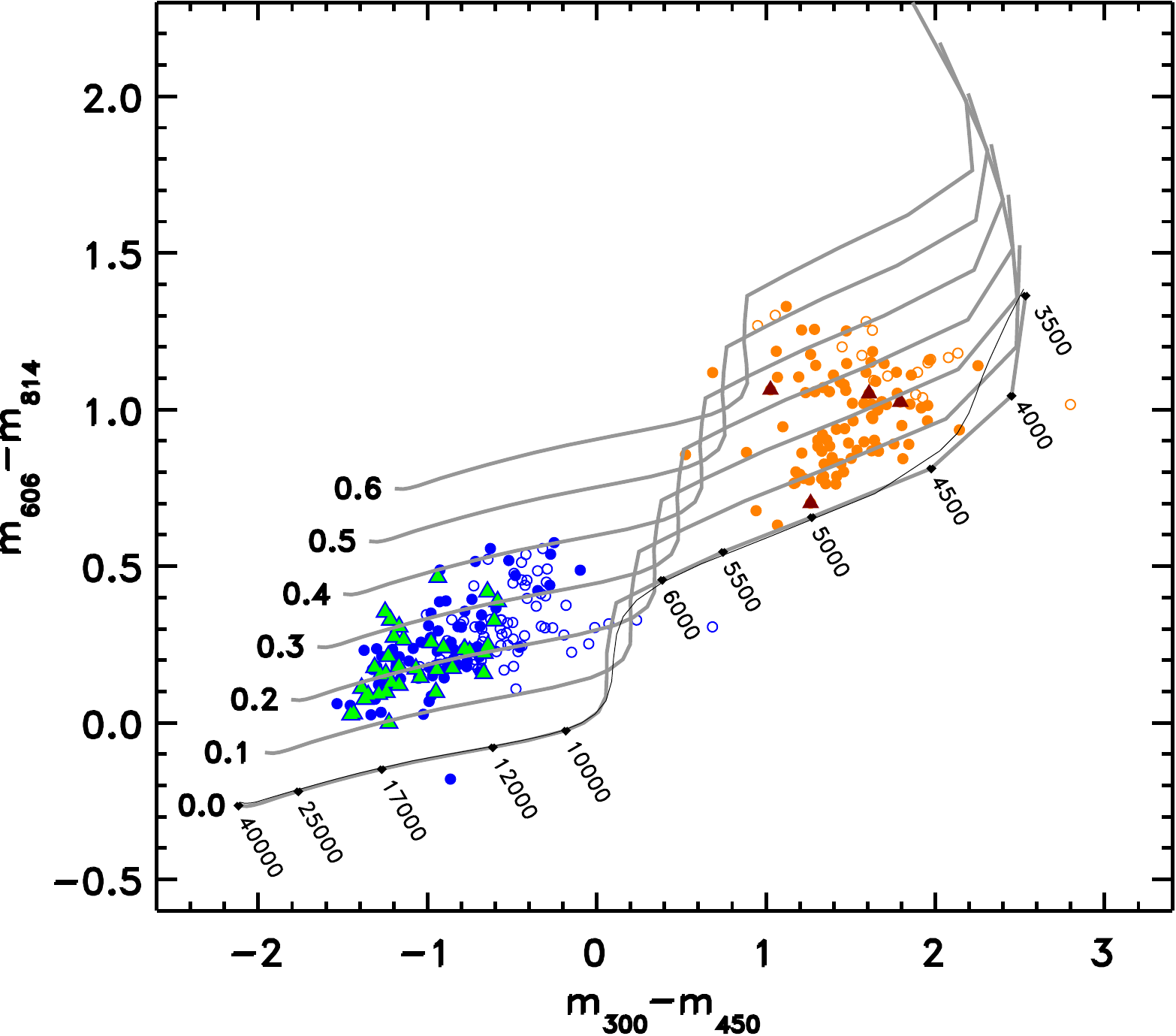}}
\caption{Colour--colour diagram for all the stars shown as thick symbols
in Figure\ref{fig10}. The symbols (and colours) are the same. The
stars shown with unfilled dots are objects with larger photometric  
uncertainty ($\delta_4 < 0.23$ instead of $\delta_4 < 0.15$). The solid
lines show the colours of the model atmospheres of Bessell et al.
(1998), to which we have applied incremental reddening as per the 
$E(B-V)$ values indicated next to each curve, according to the
extinction law that we have derived in Figure\,\ref{fig9}.}
\label{fig11}
\end{figure}

We derive the reddening towards each of the selected UMS and RC stars
from the colour--colour (CC) diagram of Figure\,\ref{fig11}, in order to
make  full use of the measurements in four bands available for all of
them. The thick ``S''-shaped solid line at the bottom of the figure
corresponds to the theoretical colours from the models of Bessell,
Castelli \& Plez (1998) for stars with gravity $\log g=4.5$, metallicity
$Z=0.006$ and effective temperature in the range 3\,500\,K --
40\,000\,K, for the specific WFPC2 photometric bands used in this work
(see also Romaniello et al. 2002). It already includes the MW
contribution to the reddening along the line of sight. For comparison,
we also show as a thin line the colours corresponding to the 4\,Myr old
isochrone of Marigo et al. (2008) for $Z=0.008$, also including the MW
contribution to the reddening. Except for the lowest temperature range,
which is outside of our region of interest, the two sets  of models
match each other very closely. 

Note that, due to the value of the surface gravity adopted for the
Bessell et al. (1998) models, the latter are in principle only suitable
to MS stars. However, Romaniello et al. (2002) have convincingly shown
that they can be effectively used also for RG stars, since the effect of
a lower surface gravity on objects with $T_\mathrm{eff} \la 5\,500$\,K
is negligible. Furthermore, the validity of these models in the WFPC2
bands used here has been specifically addressed by Romaniello et al.
(2006).

Our goal is to derive information on the interstellar reddening by 
comparing the colours of the observed stars with those of the models.
Since the ratio of the colour  excesses in different bands is set by the
extinction law that we derived in the previous sections, the
displacement of the S-curve in the CC plane must take place accordingly.
At first sight it may then seem sufficient to shift solidly the bottom 
S-curve along both axes of the CC diagram, according to the extinction
law, until the models agree with the data. However, although $R_\lambda$
is fixed by the extinction law at a given wavelength, the extinction
value $A_\lambda$ corresponding to a given $E(B-V)$, once averaged over
the bandwidth, depends on the temperature of the star. This is due to
the non negligible bandwidth of the filters causing the effective
extinction in a band to depend on the shape of the spectrum, which in
turn is a function of the star's temperature. As Romaniello et al.
(2002) and Girardi et al. (2008) have shown, this effect is more
pronounced for cold stars than for hot stars and for blue filters than
for red filters. 

This implies that reddening not only moves the S-curve to the right and
upward in the CC plane, but it also modifies its shape at redder
colours, making it steeper, i.e. more curved. This can be seen in
Figure\,\ref{fig11}, where the lowest thick grey curve corresponds to
the models of Bessell et al. (1998), to which we have already applied
the attenuation corresponding to the Galactic extinction along the line
of sight to the LMC. The other thick grey curves represent the same
models, to which we have further applied an attenuation corresponding to
incremental values of $E(B-V)$, assuming the extinction specific to this
region as derived in the previous section. In practice, since the
parametric description of Fitzpatrick \& Massa (1999) with $R_V=5.6$
agrees very well with our extinction law (see Figure\,\ref{fig9}), we
have used it to calculate the attenuation of the model atmospheres over
the entire wavelength range covered by the models, for values of $\Delta
E(B-V)$ ranging from 0 to 1 in steps of $0.01$ mag. Using {\em Synphot}
(Laidler et al. 2005), we have folded the models attenuated in this way
through the specific WFPC\,2 bands and derived the corresponding
magnitudes in the {\em HST} system. The thick solid lines in
Figure\,\ref{fig11} correspond to values of $\Delta E(B-V)$ ranging from
0 to $0.6$ in steps of $0.1$, as indicated. 

\begin{figure}
\centering
\resizebox{\hsize}{!}{\includegraphics{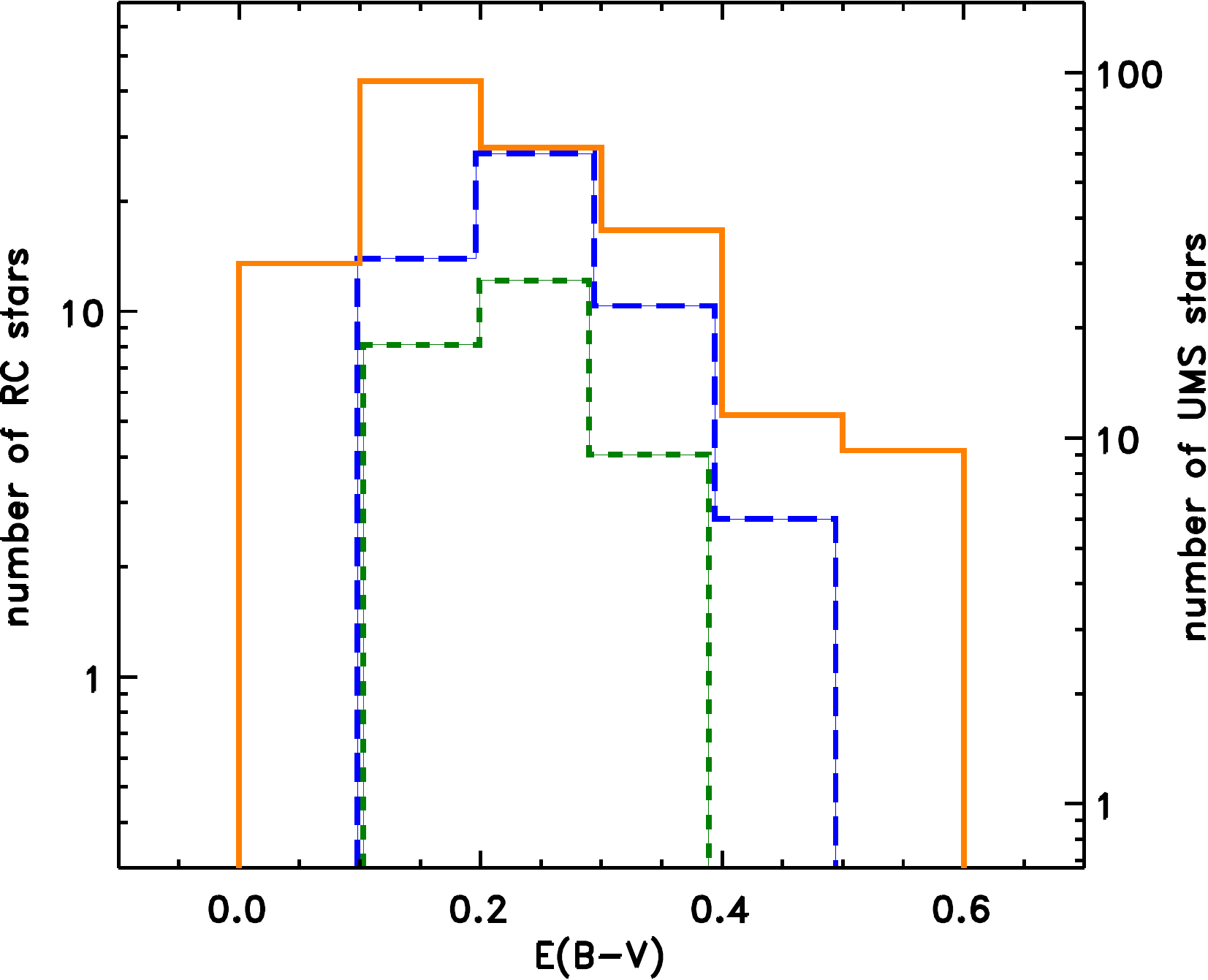}}
\caption{Histograms of the number of RC and UMS stars. The solid line
(orange in the online version) gives the number of RC stars, according
to the ordinates at left. The reddening distribution of UMS stars is
indicated by the dashed lines: the short-dashed line (green in the online
version) is for the stars in the small cluster, while the long-dashed 
line (dark blue in the online version) is for all other UMS stars in the
field. The number of UMS stars is given by the ordinates at right.}
\label{fig12}
\end{figure}

Comparing the distribution of UMS and RC stars in the figure reveals
that the former span a narrower range of $\Delta E(B-V)$ values and in
particular the UMS stars in the small cluster (triangle) have the
narrowest range. Conversely, RC stars are distributed over a wider range
of $\Delta E(B-V)$, extending both to lower and higher values. This can
be seen more easily in Figure\,\ref{fig12}, showing the histograms of
the distribution of RC stars (solid line), the UMS stars inside the
small cluster (short-dashed line) and the UMS elsewhere in the field
(long-dashed line). The spread in the $\Delta E(B-V)$ values of UMS
stars is possibly slightly overestimated, since some of it can be due to
stellar evolution. In other words, an age difference of a few Myr can
imply a significant colour difference in the CMD for UMS stars (see e.g.
Figure\,\ref{fig11}), whereas for RC stars the dominant cause of the
spread observed in the CMD is reddening, as discussed in Section\,4. 

\begin{figure}
\centering
\resizebox{\hsize}{!}{\includegraphics{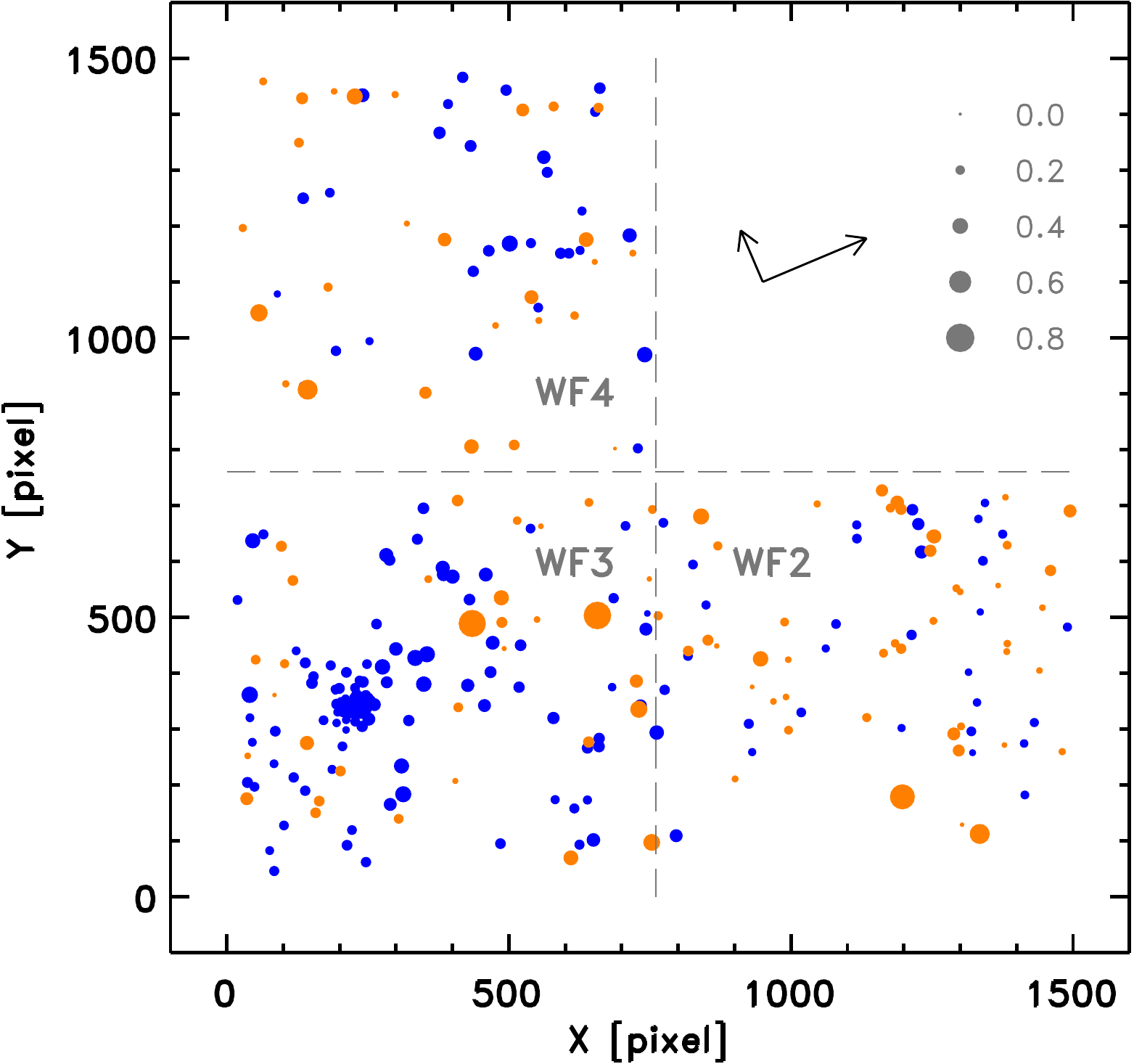}}
\caption{Map of the measured $\Delta E(B-V)$ values towards the RC stars
(orange dots) and UMS stars (blue dots) in the sample. The orientation
is the same as in Figure\,\ref{fig1} and the size of the symbols is
proportional to the $\Delta E(B-V)$ value, as per the legend. Each pixel
corresponds to $0\farcs1$.}
\label{fig13}
\end{figure}

{The spatial scale of the extinction variations is illustrated in
Figure\,\ref{fig13}, which shows a map of the $\Delta E(B-V)$ values
towards the RC (orange dots) and UMS (blue dots) in the sample. The
orientation is the same as in Figure\,\ref{fig1} and the size of the
symbols is proportional to the $\Delta E(B-V)$ value, as per the legend.
The values of the minimum, maximum, average and standard deviations of
$\Delta E(B-V)$ measured in the three WF chips are given in
Table\,\ref{tab4}, separately for RC and UMS stars. The table reveals
that the average extinction appears to not vary much in the field,
although in chip WF\,2 its value is marginally lower. The fact that RC
stars have both lower and higher extinctions than the UMS objects
suggests that they are not directly correlated with the absorbing
material and that they are distributed everywhere: within the cloud that
envelopes the UMS stars and both in front and behind it. The absorbing
material, on the other hand, is most likely associated with the UMS
stars themselves or anyhow with the molecular cloud out of which they
recently formed. It is, thus, not surprising that RC stars located
behind the far edge of the molecular cloud show higher extinction.}

\begin{table}
\centering 
\caption{Statistics on the extinction towards RC and UMS stars across
the observed field. For each of the three WF chips (see
Figure\,\ref{fig13}), the table gives the number of stars ($N$) in the RC
or UMS sample (Type) and some statistics on the extinction towards these
objects, namely the minimum (Min), maximum (Max), average (Avg) and
standard deviation (Stdev) of the measured $\Delta E(B-V)$.}
\begin{tabular}{llccccc} 
\hline
Chip & Type & $N$ & Min & Max & Avg & Stdev \\
\hline
WF\,2 & RC &  46 & $0.04$ & $0.69$ & $0.21$ & $0.13$ \\
WF\,3 & RC &  32 & $0.04$ & $0.77$ & $0.26$ & $0.17$ \\
WF\,4 & RC &  27 & $0.02$ & $0.53$ & $0.23$ & $0.12$ \\
\hline
WF\,2 &UMS &  31 & $0.12$ & $0.32$ & $0.20$ & $0.05$ \\
WF\,3 &UMS & 112 & $0.11$ & $0.41$ & $0.25$ & $0.07$ \\
WF\,4 &UMS &  30 & $0.13$ & $0.41$ & $0.25$ & $0.07$ \\
\hline      
\end{tabular}
\vspace{0.5cm}
\label{tab4}
\end{table}

\begin{figure}
\centering
\resizebox{\hsize}{!}{\includegraphics{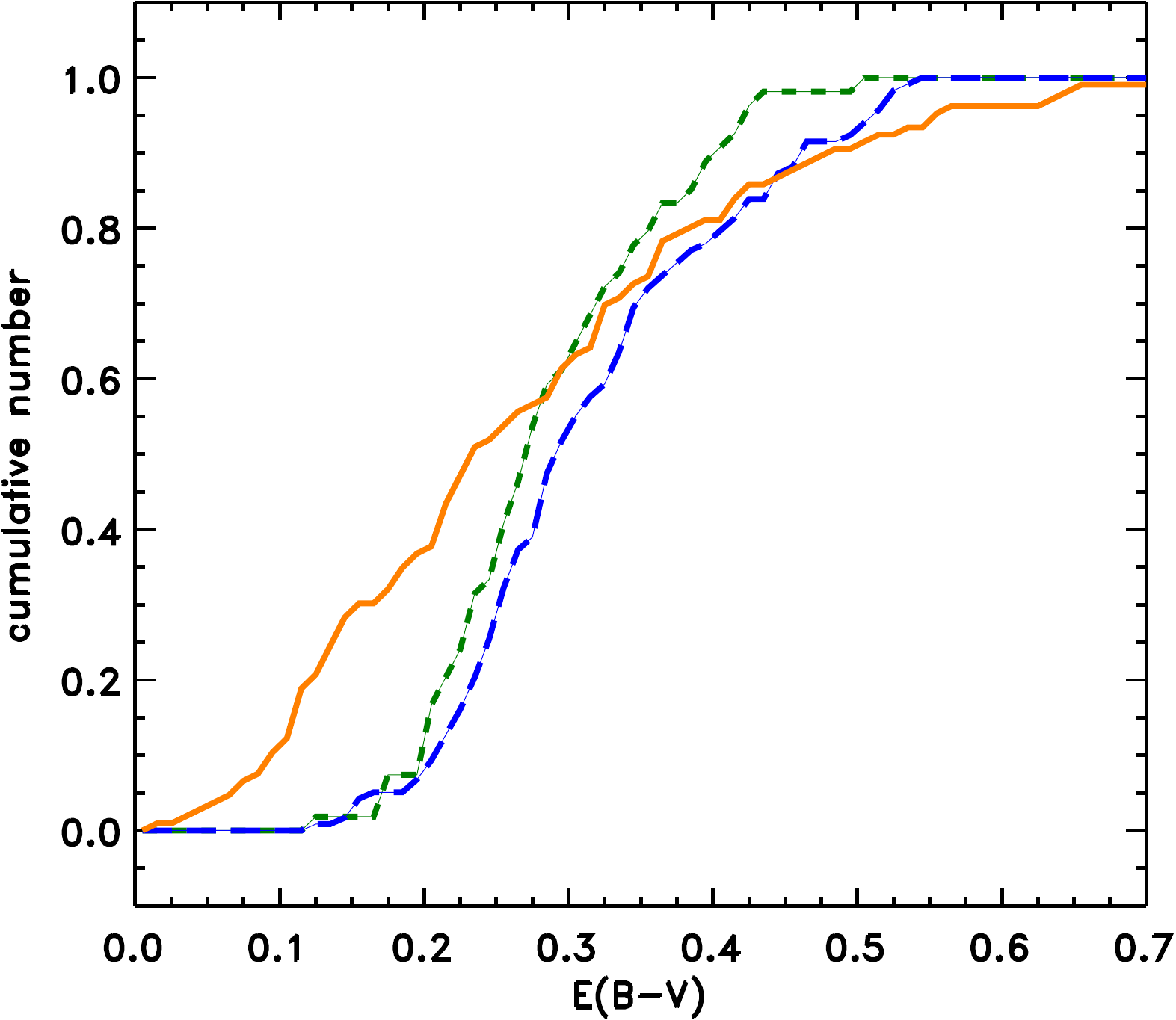}}
\caption{Cumulative distributions of the RC stars (solid line, orange in
the online version), UMS stars in the small cluster (short-dashed
line, green in the online version) and other UMS stars in the field
(long-dashed line, blue in the online version). A Kolmogorov--Smirnov
test shows that the probability that RC and UMS stars are drawn from the
same underlying distribution is very small, $P=7.2 \times 10^{-5}$.}
\label{fig14}
\end{figure}

To better quantify the difference in the reddening distribution of UMS
and RC stars we show in Figure\,\ref{fig14} the cumulative distributions
of these objects, as a function of $E(B-V)$. A Kolmogorov--Smirnov test
shows that the probability that RC and UMS stars come from the same
distribution is very small, namely $P=7.2 \times 10^{-5}$. As a sanity
check we have also compared the distribution of the UMS stars in the 
cluster (short-dashed line) and over the rest of the field (long-dashed
line), finding that they are indeed not significantly different
($P=0.57$). 

As mentioned above, Zaritsky (1999) reported that in the LMC stars with
$T_{\rm eff} > 12\,000$\,K are typically affected by larger extinction
values (several tenths of a magnitude more) than stars with $5\,500$\,K$
< T_{\rm eff} < 6\,500$\,K. Our results, therefore, appear at variance
with Zaritsky's (1999) conclusions, since Figure\,\ref{fig12} shows that
the extinction towards RC stars can be considerably larger than that
measured towards UMS objects. Note that we are confident about the 
nature of the RC objects, since we have only retained objects with 
$W_{\rm eq}(H\alpha) < 3$\,\AA\ in order to carefully exclude PMS stars
and other possible interlopers.

In fact, that the reddening is on average less severe for young stars
than for the older objects was recently found by Sabbi et al. (2013) in
the Tarantula Nebula, containing the 30\,Dor region. As pointed out in
that work, in his analysis Zaritsky (1999) only considered stars that
were detected simultaneously in all four photometric bands  $U$, $B$,
$V$ and $I$. This works as a bias against the more extinguished old
stars, since they are the first ones to drop below the detection limit
in the $U$ band. Therefore, it is possible that in Zaritsky's (1999)
study the average amount of reddening towards red giant branch stars 
might have been underestimated. 

On the other hand, it is also possible that, with a field of view of
$\sim  2\farcm7$ or $\sim 40$\,pc on a side, our observations do not
sample a region wide enough to probe a statistically representative
sample of UMS stars. For instance, the objects belonging to the small
cluster have an even more limited spead of extinction than the other UMS
objects in the field because they occupy a small volume along the line
of sight and, more in general, this might be true for the entire young
population in this field. Conversely, being older, stars like RC objects
are more uniformly distributed. Since we are probing a particular region
of the LMC, there is no certainty that our findings apply in general to
this galaxy. We plan to extend this study to a much wider area, using
existing high qualify {\em HST} observations of the central regions of 30\,Dor
(De Marchi et al. 2011b) and of a large portion of the Tarantula Nebula
(Sabbi et al. 2013). These observations will allow us to determine the
reddening distribution towards both UMS and RC stars over a region of
active star formation, covering almost 1\,\% of the entire area of the
LMC. 

\vspace*{-0.5cm}
\section{Summary and conclusions}

We have studied the properties of the interstellar extinction in a field
of about $2\farcm7$ on a side located about $6\arcmin$ SW of 30\,Dor in 
the LMC. The observations with with the WFPC\,2 on board the {\em HST} in the
$U$, $B$, $V$, $I$ and $H\alpha$ bands  show the presence of patchy
extinction in this field. In particular, the CMD reveals an elongated
stellar sequence, running almost parallel to the MS, which is in
reality made up of RC stars spread across the CMD because of the
considerable and uneven levels of extinction across this region. This 
allows us to derive in a quantitative way both the extinction law 
$R_\lambda$ and the values of the absolute extinction $A_\lambda / A_V$
towards about 100 objects, thereby setting statistically significant
constraints on the properties of the extinction in this area. The main
results of this work can be summarised as follows.

\begin{enumerate}

\item  

Using theoretical CMDs of the same type as those presented by Girardi \&
Salaris (2001) and Salaris \& Girardi (2002), but computed for the
specific WFPC2 bands of our observations, and assuming a single stellar
population, we have studied the expected behaviour of the mean RC as a
function of age (from $1.4$\,Gyr to $3.0$\,Gyr) and metallicity (from
$1/20\,Z_\odot$ to $Z_\odot$). We show that metallicity, age and
reddening affect the position of RC stars in different ways in the CMDs
defined by our bands. 

\item

Our analysis shows that the magnitude of the RC is not affected
considerably by the age of the stars, except at very low metallicity. In
bands longwards of $\sim 4\,000$\,\AA\ an increase of metallicity
would move the RC along directions similar to the reddening vector.
However this effect is small because even a change of metallicity by a
factor of two would mimic a quite modest amount of reddening. This
limited sensitivity of the RC to age and metallicity makes it easier to
estimate the amount of reddening when there is considerable interstellar
extinction ($A_V > 1$), since in this case the magnitude and colour
displacement of the RC in the CMD due to extinction dominate over all
uncertainties on metallicity and age. 

\item 

To identify bona-fide RC stars, we have compared our observations in all
bands with the theoretical CMDs for the metallicity range applicable to
the LMC ($0.004 < Z/Z_\odot < 0.008$), taking into account the known
distance modulus and intervening MW reddening. This allows us to define
the region of the CMDs where reddening can place RC stars, finding 170
objects inside it. With an iterative procedure, we have reduced this
number to a total of 107 bona-fide RC stars simultaneously classified as
such in  all CMDs and with combined photometric uncertainty $\delta_4 <
0.23$\,mag. We have conservatively removed from this sample two
bona-fide PMS stars, with  H$\alpha$ excess emission in the range
10\,\AA\ $< W_{\rm eq}(H\alpha) < 30$\,\AA, and additional ten
objects with 3\,\AA\ $< W_{\rm eq}(H\alpha) < 10$\,\AA\ that might also
be PMS stars. 

\item 

In each CMD, the best linear fit to the distribution of the bona-fide RC
stars provides the absolute extinction and the ratio $R$ between 
absolute and selective extinction in the specific bands of our
observations. Through interpolation at the wavelengths of the Johnson
$B$ and $V$ bands, we have derived the extinction curve in the form
$R_\lambda  \equiv A_\lambda / E(B-V)$, in the range $\sim 3\,000 -
8\,000$\,\AA. Adopting the parametrization of Cardelli et al. (1989) and 
Fitzpatrick \& Massa (1999), in this wavelength range our extinction law
is consistent with a value of $R_V=5.6$ with an uncertainty of $0.3$.

\item

Compared with the values reported in the diffuse Galactic interstellar 
medium, where on average $R_V = 3.1$ (Cardelli et al. 1989), and with
those towards 19 lines of sight in the LMC explored by Gordon et al.
(2003), where on average $R_V=3.4$, our larger $R_V$ value indicates a
considerably flatter extinction law, which in turn implies that in this
region larger grains dominate. Since larger $R_V$ values are usually
found in denser regions, it is possible that the specific area that we
have studied is simply denser than those sampled in the LMC so far. More
in general, it could be the result of systematically different 
conditions of the environments in the LMC as compared to the MW. 

\item   

Having derived the extinction law in this region, we have compared the
extinction towards individual RC and UMS stars in order to study the
relative distribution of stars and dust in this field. We find that UMS
objects span a narrower range of $E(B-V)$ values than RC stars, at 
variance with the conclusion of Zaritsky (1999) that in the LMC stars
with $T_{\rm eff} < 12\,000$\,K are typically affected by larger
extinction than stars with $5\,500$\,K $< T_{\rm eff} < 6\,500$\,K.
While it is plausible that in Zaritsky's (1999) study the average amount
of reddening towards red giants might have been underestimated (see
Sabbi et al. 2013), it is also possible that we are probing a peculiar
region of the LMC. 

\end{enumerate}

In summary, in this work we have shown how to use multi-band {\em HST}
photometry of a resolved stellar population to efficiently derive
simultaneously the extinction law and the absolute value of the
extinction towards a large number of RC stars in a typical LMC field.
When the distance to the sources and their metallicity range are known,
which is the case of any field in the MC, the average location of the RC
in a CMD can be used to derive the absolute extinction towards its
individual members by comparison with theoretical models. We have shown
how multi-band photometry, and particularly the inclusion of the $U$
band, can be used to lift the degeneracy introduced by uncertainties on
the age and metallicity. The results are statistically significant
because several dozen RC stars are present in the average MC field.

Furthermore, when the amount of the extinction is high and variable
across the field, like in the region that we have studied, also the
extinction law can be accurately derived, in absolute terms. In general,
when there is measurable extinction, even without an appreciable
spread, one can still determine the extinction law from the displacement
between expected and actual RC locations, as recently done for instance
by Nataf et al. (2013; see also Udalski 2003 and Sumi 2004, and references 
therein) using the OGLE observations of the Galactic
bulge. However, the determination in this case is less reliable, because
the lack of a clear reddening track makes the identification of spurious
systematic effects much harder.

The advantage of our method is that the RC stars that we consider in our
study are all at the same distance, to within 1\,\%, that they have very
similar intrinsic physical properties in all bands, and that we can
derive the absolute value of the extinction without having to extend the
observations to infrared bands. So far our investigation using this
method has covered a single field near 30\,Dor, in which we find a much
shallower extinction law than so far reported for the LMC, corresponding
to $R_V=5.6 \pm 0.3$. We plan to apply this method to existing {\em HST}
observations of the 30\,Dor cluster and of the Tarantula nebula at
large, in order to investigate whether this extinction law is
characteristic of other regions of intense star formation in the MC.

\section*{Acknowledgements}

We are grateful to Vera Kozhurina--Platais for her assistance with the
analysis of the data, and to Ian Howarth, the referee, for insightful
comments that have helped us to improve the presentation of this work.
NP acknowledges partial support by STScI--DDRF grant D0001.82435.

\end{document}